\documentclass[aps,notitlepage,preprintnumbers,onecolumn,amsmath,amssymb,floatfix,superscriptaddress,pra]{revtex4-1}

\pdfoutput=1 
\usepackage[utf8]{inputenc}

\usepackage{amsmath}
\usepackage{amsfonts}
\usepackage{braket}
\usepackage{graphicx}
\usepackage[usenames,dvipsnames,svgnames,table]{xcolor}

\usepackage[colorlinks=true,linkcolor=blue, citecolor=blue, bookmarks]{hyperref}

\def\Tr{\mbox{Tr}\,}

\newcommand{\be}{\begin{equation}}
\newcommand{\ee}{\end{equation}}
\newcommand{\bea}{\begin{eqnarray}}
\newcommand{\eea}{\end{eqnarray}}

\begin{document}
\title{Correlation and entanglement spreading in nested spin chains}
\author{Ranjan Modak} 
\affiliation{SISSA and INFN, via Bonomea 265, 34136 Trieste, Italy}
\author{Lorenzo Piroli}
\affiliation{Max-Planck-Institut f\"ur Quantenoptik, Hans-Kopfermann-Str. 1, 85748 Garching, Germany}
\author{Pasquale Calabrese}
\affiliation{SISSA and INFN, via Bonomea 265, 34136 Trieste, Italy}
\affiliation{International Centre for Theoretical Physics (ICTP), Strada Costiera 11, 34151 Trieste, Italy}

\begin{abstract}
	
The past few years have witnessed the development of a comprehensive theory to describe integrable systems out of equilibrium, in which the Bethe ansatz formalism has been tailored to address specific problems arising in this context. While most of the work initially focused on the study of prototypical models such as the well-known Heisenberg chain, many theoretical results have been recently extended to a class of more complicated \emph{nested} integrable systems, displaying different species of quasiparticles. Still, in the simplest context of quantum quenches, the vast majority of theoretical predictions have been numerically verified only in systems with an elementary Bethe ansatz description. In this work, we fill this gap and present a direct numerical test of some results presented in the recent literature for nested systems, focusing in particular on the Lai-Sutherland model. Using time-dependent density matrix renormalization group and exact diagonalization methods, we compute the spreading of both correlation functions and entanglement entropy after a quench from a simple class of product initial states. This allows us to test the validity of the nested version of a conjectured formula, based on the quasiparticle picture, for the growth of the entanglement entropy, and the Bethe ansatz predictions for the ``light-cone'' velocity of correlation functions. 
\end{abstract}
\maketitle

\section{Introduction}
\label{sec:intro}

After more than a decade of intense investigations, the theory of many-body integrable systems out of equilibrium is arguably entering a more mature stage, as several experimentally feasible protocols are now falling under the reach of our theoretical understanding \cite{CaEM16,KiWW06,LaGS16,LEGR15,SBDD19,FSEH13}. In this respect, the study of the quantum quench \cite{cc-06}, in which a well-defined initial state is left to evolve unitarily according to some known Hamiltonian, has played a major role for the development of the theory. Indeed, besides offering a conceptually simple framework to investigate questions of fundamental interest, the need to overcome the significant computational challenges of this problem led to the introduction of powerful methods \cite{EsFa16,Caux16,IMPZ16}, which now represent significant milestones in the field.

One of the most important ones is arguably the Quench Action approach introduced in Ref.~\cite{CaEs13}, and first applied in Ref.~\cite{DWBC14} to a genuinely interacting model (see Ref.~\cite{Caux16} for a review). This method provided an intuitive picture to describe the local properties of the system at large times after a quench: according to the Quench Action approach, the unitary dynamics brings the system to a stationary state whose local properties can be computed from the knowledge of a single representative eigenstate of the Hamiltonian. Importantly, the latter can be characterized in terms of the corresponding quasimomentum (or rapidity) distribution functions of the stable quasiparticles; in fact, it was later realized \cite{BWFD14,PMWK14,IDWC15,PoVW17,IlQC17,IQDB16,PiVC16,PVCR17} that this characterization is equivalent to that of the generalized Gibbs ensemble (GGE) \cite{RDYO07,ViRi16,CaEF11,FaEs13,Pozs13}, a statistical ensemble which generalizes the Gibbs density matrix by taking into account all the local and quasilocal conservation laws of the Hamiltonian~\cite{Pros11,IlMP15,PiVe16,DeCD17,PrIl13}.

The description of the post-quench steady state in terms of the quasiparticle rapidity distribution functions is particularly convenient, as they are directly related to several aspects of the post-quench dynamics. For instance, the ``light-cone'' velocity of correlation functions was argued to coincide with that of the  fastest quasiparticle excitation built on top of the stationary state \cite{BoEL14}, while exact formulas have been derived to compute the post-quench steady value of local correlators as a function of these rapidity distributions \cite{Pozs11,MePo14,Pozs17,BaPi18,BePC16,NeSm13}.  Furthermore, as a notable result, it was shown in Refs.~\cite{AlCa17,AlCa18} that they also give us access to the exact evolution of the entanglement after the quench (in the scaling limit of asymptotically large times and spin blocks), as the knowledge of the occupation numbers and dispersion relations of the quasiparticles complements the standard picture of Ref.~\cite{CaCa05} introduced in the context of conformal field theory.

Unfortunately, the actual computation of the rapidity distributions of the quasiparticles remains in general a difficult task. Indeed, the prescriptions of Ref.~\cite{CaEs13} to compute them require, as an input, the overlaps between the initial state and the eigenstates of the post-quench Hamiltonian, which have proven to be very hard to obtain \cite{KoPo12,Pozs14,cd-14,PiCa14,Broc14,LeKZ15,LeKM16,HoST16,BrSt17,Pozs18,LeKL18,HoKT19}. In order to get around this problem, two different approaches were later introduced: the first one is the ``string-charge duality'' method \cite{IQDB16,IDWC15}, based on the idea that the post-quench representative eigenstate could be uniquely fixed from the expectation value of a complete set of quasi-local conserved operators on the initial state. While this method allows us to study quenches from rather general initial states, it is limited to the models for which the latter are known; in particular, so far it could be applied only to XXZ Heisenberg chains \cite{IQDB16,PVCR17,PiVC16}. The second method is the Quantum Transfer Matrix approach introduced in Refs.~\cite{PiPV17,PiPV17_II} (see also \cite{PiPV18_losch}), which derives the post-quench rapidity distribution functions based on some underlying mathematical structures of integrability. This approach is, by construction, limited to a special class of ``integrable'' initial states \cite{PiPV17_II,PoPV18}, whose definition is inspired by classical works in integrable quantum field theory \cite{Ghos94} (see also Refs.~\cite{Delf14,Schu15}); still it was immediately realized it could be applied quite generally to different models, relying exclusively on defining properties of integrability \cite{PiPV17_II}.

Given the significant technical challenges, these important developments were initially restricted to classes of models solvable through an elementary (as opposed to \emph{nested}) Bethe ansatz (with a notable exception in the context of relativistic quantum field theory \cite{BeSE14}). Still, nested integrable systems are particularly interesting as they contain different species of interacting particles \cite{efgk-05}, and include experimentally relevant examples such as, most prominently, multi-component Fermi and Bose gases, as already realized in several pioneering cold atomic experiments \cite{BlDZ08,guan2013,PMCL14,exp2}. As a recent development, a series of works was able to extend several of the aforementioned results to the case of nested systems, focusing on the simplest example of the Lai-Sutherland chain \cite{lai-74,sutherland-75}, a special case of the Perk-Schultz model~\cite{PeSc81,Schu83}. In particular, the first application of the Quench Action method in this system appeared in Ref.~\cite{MBPC17}, where some of us analyzed the quench from a very special initial state, whose overlaps with the eigenstates of the Hamiltonian were computed in Ref.~\cite{LeKM16}. Later, the Quantum Transfer Matrix approach was generalized in Refs.~\cite{PVCP18_I, PVCP18_II} to the nested models, finally allowing us to study the post-quench dynamics for an entire class of integrable initial states. Note that the special state considered in Ref.~\cite{LeKM16} was also shown to be a particular example in this class \cite{PiPV17_II,PoPV18}.

The theory that has been developing with the introduction of the Quench Action approach relies on several assumptions, although physically very sound, and extensive numerical tests have been necessary to establish its validity beyond reasonable doubt. In particular, the predictions of the theory for the asymptotic values of local correlations have been tested against tDMRG \cite{WhFe04,DKSV04} and iTEBD \cite{Vida07} computations \cite{FCEC14,PVCR17,PMWK14}, as well as by means of exact diagonalization techniques and numerical-linked cluster expansions \cite{BWFD14, PVCR17}. Analogously, the evolution of the entanglement has been tested extensively against tDMRG simulations in Refs.~\cite{AlCa17,AlCa18}, as well as other entanglement measures \cite{AlCal-b,AlCal-c}. However, all of these works focused on the prototypical XXZ Heisenberg chain, while no direct verification was ever performed for the case of nested systems (with the only exception of Fermi-Hubbard models with infinite onsite repulsion~\cite{BeTC17,BeTC18,ZhVR18,ZhVR19}, which is however simpler, since it can be mapped to a free model). Evidently, it is very important to corroborate at least some of the recent results obtained in nested systems. This is the aim of this work, where we focus on the Lai-Sutherland chain, and compute numerically different quantities after the quench from the family of integrable initial states of Refs.~\cite{PVCP18_I,PVCP18_II}. This allows us to test directly several analytic predictions formulated in the recent literature for a truly interacting nested intergrable model, thus corroborating the existing theoretical framework for the study of quantum quenches in nested models.

The paper is organized as follows. In Sec.~\ref{sec:setup} we introduce the Lai-Sutherland model, and discuss the class of initial states which will be considered in this work. Our numerical results for the evolution of the entanglement are reported in Sec.~\ref{sec:entanglement_dynamics}, together with a comparison against analytic predictions available in the literature. Sec.~\ref{sec:correlations} is devoted to the analysis of the correlation functions after the quench, while our conclusions are consigned to Sec.~\ref{sec:conclusions}. Finally, technical details on our numerical computations are reported in Appendix~\ref{sec:numerics}.

\section{The model and the initial states} \label{sec:setup}

We consider the Lai-Sutherland model~\cite{lai-74,sutherland-75}, which is described by the Hamiltonian
\begin{eqnarray}
H_L=\sum_{j=1}^{L-1}\textbf{s}_j.\textbf{s}_{j+1}+(\textbf{s}_j.\textbf{s}_{j+1})^{2} -2L\,,
\label{eq:hamiltonian}
\end{eqnarray}
where $\textbf{s}_j=(s^{x}_j,s^y_j,s^z_j)$. The spin-$1$ operators $s^{a}_{j}$ act on the local Hilbert space $h_j\simeq \mathbb{C}^3$, and are given by the standard three-dimensional representation of the
$SU(2)$ generators; in particular, introducing the local basis $\ket{1}=(1,0,0)$, $\ket{2}=(0,1,0)$, $\ket{3}=(0,0,1)$, they read
\begin{gather}
 s^x=\frac{1}{\sqrt{2}}\begin{bmatrix} 0 & 1 & 0 \\ 1 & 0 & 1 \\ 0 & 1 & 0  \end{bmatrix}\,,\quad 
 s^y=\frac{1}{\sqrt{2}}\begin{bmatrix} 0 & -i & 0 \\ i & 0 & -i \\ 0 & i & 0 \end{bmatrix}\,\quad  
 s^z=\begin{bmatrix} 1 & 0 & 0 \\ 0 & 0 & 0 \\ 0 & 0 & -1 \end{bmatrix}\,.
\end{gather}

The calculations presented in this work will be exclusively numerical. However, since they will be compared to analytic predictions, it is useful to recall a few basic aspects of the theoretical tools used to analyze the Hamiltonian~\eqref{eq:hamiltonian}, which provides one of the simplest models which can be solved via the nested Bethe ansatz method \cite{ efgk-05,kr-81,johannesson-86,johannesson2-86}. Specifically, by means of the latter it is possible to show that the eigenstates of \eqref{eq:hamiltonian} are parametrized by \emph{two} distinct sets of quasi-momenta (or rapidities) $\{k_{j}\}_{j=1}^N$, $\{\lambda_{j}\}_{j=1}^M$, satisfying an appropriate set of quantization conditions (the nested Bethe equations)
\bea
\left(\frac{k_{j}+i/2}{k_{j}-i/2}\right)^{L}=\prod_{\scriptstyle p=1\atop \scriptstyle p\ne j}^{N} \frac{k_j- k_p+ i}{k_j- k_p- i} \prod_{\ell = 1}^{M} \frac{\lambda_\ell- k_j+ i/2}{\lambda_\ell- k_j- i/2}\,, \quad j=1,\ldots,N\,,\label{eq:bethe_equations1}\\
1=\prod_{j=1}^N \frac{k_j-\lambda_\ell- i/2}{k_j-\lambda_\ell+ i/2}\prod_{\scriptstyle m=1\atop \scriptstyle m\neq \ell}^M  \frac{\lambda_\ell-\lambda_m- i}{\lambda_\ell-\lambda_m+ i}\,, \quad \ell=1,\ldots,M\,.
\label{eq:bethe_equations2}
\eea
The energy eigenvalues can then be written as 
\be
E=-\sum_{j=1}^{N}1/(k_j^2+1/4)\,.
\label{eq:energy}
\ee
Physically, the sets $\{k_{j}\}_{j=1}^N$ and $\{\lambda_{j}\}_{j=1}^M$ correspond to two distinct types, or species, of quasiparticles. This situation is different from that of ordinary Bethe-ansatz solvable models (e.g. the well-know Heisenberg chain \cite{kbi-93}), where the eigenstates of the Hamiltonian are parametrized by a single set of rapidities. We note that the existence of more than one species of quasiparticles is not just a mathematical fact, but have implications on observable quantities, even in out-of-equilibrium protocols~\cite{MBPC19}.

The thermodynamic description of the model, defined by $L,N,M\to \infty$ keeping the ratios $D_1=N/L$ and $D_2=M/L$ constant, is analogous to the case of ordinary Bethe ansatz \cite{takahashi_book}. In particular, in this limit the rapidities arrange themselves in the complex plane according to specific patterns called strings, which correspond to bound-states of the quasiparticles; in each string the rapidities are parametrized as $ k^{n,\ell}_{\alpha}=k^{n}_{\alpha}+i\left[(n+1)/2-\ell\right]+\delta_{1,\alpha}^{n,\ell}$, 
$\lambda^{n,\ell}_{\alpha}=\lambda^{n}_{\alpha}+i\left[(n+1)/2-\ell\right]+\delta_{2,\alpha}^{n,\ell}$, where $\ell=1,\ldots n$. Here the real numbers $k^{n}_{\alpha}$, $\lambda^{n}_{\alpha}\in (-\infty,+\infty)$ are the string centers, which can be interpreted as the quasimomenta of the bound-states, while $\delta_{r,\alpha}^{n,\ell}$ are negligibly small deviations; finally, $n$ is the length of the string (namely, the number of quasiparticles in the bound-state).

In the thermodynamic limit the string centers for the two species become continuous variables on the real line, distributed according to rapidity distribution functions $\rho^{(1)}_n(k)$ and $\rho^{(2)}_n(\lambda)$.  One also needs to introduce the functions $ \rho^{(1)}_{h,n}(k)$ and $ \rho^{(2)}_{h,n}(\lambda)$ describing the distribution of ``holes'', which are the available rapidities for which there is no quasiparticle. These functions are related by the following thermodynamic version of the Bethe equations (see e.g. Ref.~\cite{MBPC17})
\bea
\rho_{t,n}^{(1)}(\lambda)&=&a_n(\lambda)-\sum_{m=1}^{\infty}\left(a_{n,m}\ast\rho_m^{(1)}\right)(\lambda)+\sum_{m=1}^{\infty}\left(b_{n,m}\ast\rho_m^{(2)}\right)(\lambda)\,,\label{eq:TBAexplicit1}\\
\rho_{t,n}^{(2)}(\lambda)&=&-\sum_{m=1}^{\infty}\left(a_{n,m}\ast\rho_m^{(2)}\right)(\lambda)+\sum_{m=1}^{\infty}\left(b_{n,m}\ast\rho_m^{(1)}\right)(\lambda)\,.\label{eq:TBAexplicit2}
\eea
Here we employed the standard definition $\rho^{(r)}_{t,n}(k)=\rho^{(r)}_{n}(k)+\rho^{(r)}_{h,n}(k)$,  together with $\left(f\ast g\right)(\lambda)=\int_{-\infty}^{\infty}{\rm d}\mu f(\lambda-\mu)g(\mu)$, and
\bea
 a_{n,m}(\lambda)&=&(1-\delta_{nm})a_{|n-m|}(\lambda)+2a_{|n-m|+2}(\lambda)+\ldots
+2a_{n+m-2}(\lambda)+a_{n+m}(\lambda)\,,\label{eq:a_mn}\\
 b_{n,m}(\lambda)&=&a_{|n-m|+1}(\lambda)+a_{|n-m|+3}(\lambda)+\ldots+a_{n+m-1}(\lambda)\,,
\label{eq:b_mn}
\eea
where $a_{n}(\lambda)=n/[2\pi(\lambda^2+n^2/4)]$. Finally, from the quasiparticle distribution functions several quantities can be computed directly. Among these, the (dressed) velocities of the quasiparticles (which are once again parametrized by the rapidities $\lambda$) are obtained through the system of integral equations
\be
\begin{aligned} \rho_{t, n}^{(2)}(\lambda) v_{n}^{(2)}(\lambda) &=\sum_{k}\left(b_{n, k} * v_{k}^{(1)} \rho_{k}^{(1)}\right)(\lambda)-\sum_{k}\left(a_{n, k} * v_{k}^{(2)} \rho_{k}^{(2)}\right)(\lambda)\,, \\ 
\rho_{t, n}^{(1)}(\lambda) v_{n}^{(1)}(\lambda) &=\frac{1}{2 \pi} \varepsilon_{n}^{\prime}(\lambda)-\sum_{k}\left(a_{n, k} * v_{k}^{(1)} \rho_{k}^{(1)}\right)(\lambda)+\sum_{k}\left(b_{n, k} * v_{k}^{(2)} \rho_{k}^{(2)}\right)(\lambda)\,.
\label{eq:velocities}
\end{aligned}
\ee
Here, $v^{(1)}_n(\lambda)$ and $v^{(2)}_n(\lambda)$ are the dressed velocities of the $n$-quasiparticle bound states of rapidity $\lambda$, for the first and second species. 

\subsection{The initial states}
\label{sec:initial_states}

As we have mentioned in Sec.~\ref{sec:intro}, until recently no initial state was known in nested systems for which an analytical study of the quench dynamics could be carried out. As a first piece of progress, in Refs.~\cite{LeKM16,LeKL18} a family of matrix product states (MPSs) \cite{PVWC06} with increasing bond dimension $\xi$ was found, for which the exact overlaps with the eigenstates of the Lai-Sutherland Hamiltonian could be worked out. While the quench dynamics from the simplest case $\xi=2$ could be analyzed in Ref.~\cite{MBPC17} by means of the Quench Action method, these states have large initial entanglement, which makes them not convenient for numerical simulations. Subsequent studies clarified the status of the special states found in Refs.~\cite{LeKM16,LeKL18} as belonging to a broader class of ``integrable'' initial states \cite{PiPV17_II,PoPV18}. Furthermore, in Refs.~\cite{PVCP18_I,PVCP18_II} additional initial product states in this class were found, for which the quasiparticle rapidity distribution functions could be derived exactly. From the numerical point of view, these states are particularly convenient: on the one hand, they display small entanglement at short times; on the other hand, they are parametrized by continuous variables, which offer the possibility of a more complete analysis.

The class of integrable states found in Refs.~\cite{PVCP18_I,PVCP18_II} read
\begin{eqnarray}
|\Psi_0\rangle=|\psi_{0}\rangle_{1,2}\otimes|\psi_{0}\rangle_{3,4}\otimes....\otimes |\psi_{0}\rangle_{L-1, L}\,,
\label{eq:initial_state}
\end{eqnarray}
with the two-site block 
\begin{equation}
 |\psi_0\rangle=c_{11}|1,1\rangle+c_{22}|2,2\rangle+c_{33}|3,3\rangle+c_{12}(|1,2\rangle+|2,1\rangle)+c_{13}(|1,3\rangle+|3,1\rangle)+c_{23}(|2,3\rangle+|3,2\rangle).
\end{equation}
Here $c_{ij}$ are arbitrary complex numbers. Note that, as it was shown in Ref.~\cite{PVCP18_I}, each state in this class can be rotated via a global $SU(3)$ transformation to a product state with a two-site block of the form
\begin{equation}
 |\psi_0\rangle=c_{11}|1,1\rangle+c_{22}|2,2\rangle+c_{33}|3,3\rangle\,,
 \label{eq:diagonal_block}
\end{equation}
with the additional restriction $|c_{11}|\geq|c_{22}|\geq|c_{33}|$. In the rest of this work, we will study the quench dynamics from the initial states \eqref{eq:initial_state}, with two-site blocks given in \eqref{eq:diagonal_block}.

Before leaving this section, we recall an important property of these states: their overlaps with an eigenstate of the Hamiltonian is non vanishing only if the latter is labeled by sets of rapidities $\{k_j\}$, $\{\lambda_j\}$ which are parity invariant, namely $\braket{\{k_j\},\{\lambda_j\}|\Psi_0}\neq 0$ implies $\{k_j\}=\{-k_j\}$, $\{\lambda_j\}=\{-\lambda_j\}$. This property has been chosen as the very definition of integrability in Refs.~\cite{PiPV17_II,PoPV18}: there, an initial MPS with finite bond dimension was called integrable if this condition is fulfilled. As we will see in the next section, this property has also consequences on the description of the post-quench entanglement dynamics.

\section{The entanglement dynamics}
\label{sec:entanglement_dynamics}

We start our analysis with the entanglement dynamics after the quench from an integrable initial state. Specifically, we consider a subsystem $A$ of length $l$, and compute numerically the evolution of the entanglement between $A$ and the rest of the system 
$B$, as given by the R\`enyi entropy
\begin{equation}
S^{\alpha}(t)=\frac{1}{1-\alpha}\ln\Tr\rho^{\alpha}_A(t)\,,
\label{eq:Renyie}
\end{equation}
where $\alpha$ is an arbitrary positive real number. Here $\rho_{A}(t)$ is the time-evolved reduced density matrix of the subsystem $A$, i.e. $\rho_{A}=\Tr_{B}|\Psi(t)\rangle\langle\Psi(t)|$, where $|\Psi(t)\rangle=e^{-iHt}|\Psi_0\rangle$. We will be particularly interested in the limit $\alpha\to 1$, which yields the well-known von Neumann entropy $S_V(t)=-{\rm tr}\left[\rho_A(t)\log \rho_A(t)\right]$.

As we have discussed in Sec.~\ref{sec:intro}, an important result in the recent literature has been the discovery in Refs.~\cite{AlCa17,AlCa18} of a formula to compute the evolution of the von Neumann entanglement entropy in the scaling limit $t,l\to\infty$, keeping the ratio $l/t$ fixed. Such formula was derived in Ref.~\cite{AlCa17} based on an intuitive quasiparticle picture, which goes along the following lines. One interprets the quench as a process generating everywhere and homogeneously uncorrelated pairs of entangled quasiparticles with opposite momenta. In this picture, two regions may be entangled only if they share at least a pair of quasiparticles emitted from an arbitrary initial point. Accordingly, the total entanglement entropy between a region $A$ and its complement $B$ is proportional to the number of pairs with one quasiparticle in $A$ and the other in $B$. Assuming there exists only one type of quasiparticles, this semiclassical description immediately gives us the following expression for the von Neumann entanglement entropy \cite{AlCa17}
\begin{equation}
S_{l}(t)\propto 2t\int_{2|v|t<l}d\lambda |v(\lambda)|s(\lambda)+l\int_{2|v|t>{l}}d\lambda s(\lambda),
\label{eq:intuitive_entanglement}
\end{equation}
where $s(\lambda)$ is the contribution to the entanglement carried by the pair of quasiparticles with rapidity $\pm \lambda$, while $|v(\lambda)|$ is the absolute value of their (opposite) velocities. In Ref.~\cite{AlCa17} it was argued that the first could be identified with the Yang-Yang entropy density \cite{YaYa69} while the second could be obtained from the dressed velocity of the quasiparticles as computed from the post-quench rapidity distribution functions \cite{BoEL14}. A straightforward generalization of this picture to the case where the quasiparticles can form bound states was employed in Ref.~\cite{AlCa17} to obtain quantitative predictions for the $XXZ$ Heisenberg chain. 
The same ideas can be also extended to inhomogeneous situations \cite{Al-18,abf-19,bertini-2018a} within the generalized hydrodynamics approach \cite{CaDY16,BCDF16}.

The analytic formula found in Ref.~\cite{AlCa17} is now widely believed to be exact for the class of integrable initial states: on the one hand, based on their defining property, the application of the quasiparticle picture is particularly intuitive and natural for these states; on the other hand, the formula for the entanglement growth was extensively tested against tDMRG simulations in Refs.~\cite{AlCa18,AlCa17} and very good agreement was always found for integrable quenches. Conversely, it is expected that these results should be modified for initial states with a more complicated structure of the overlaps, as it was found in Refs.~\cite{BeTC18,bc-18} for the case of free models.  
It is important to stress that a similar formula for the time evolution of R\'enyi entropies after a quench to an interacting integrable models is not yet known \cite{AlCal-b}.

In the case of nested systems, it is natural to assume that a similar picture holds for integrable initial states, and that the quasiparticles of each species and their bound states carry an independent contribution to the entanglement. In the case of the Lai-Sutherland model, this leads to the formula \cite{MBPC17,PVCP18_I}
\begin{equation}
\lim_{\substack{t,l\to\infty\\ l/t\ {\rm fixed}}}S_l(t)/l=(S^{(1)}_l+S^{(2)}_l)/l\,,
\label{eq:entanglemententropy}
\end{equation}
where
\be
S^{(r)}_l/l=\sum_{n=1}^{\infty}\int\!\!{\rm d}\lambda\,\, s_{n}^{(r)}(\lambda)\left\{2 \frac{t}{l} |v_{n}^{(r)}(\lambda)|\,\theta_{\rm H}\left(\frac{l}{t}-{2|v_{n}^{(r)}(\lambda)|}\right)+\, \theta_{\rm H}\left({2|v_{n}^{(r)}(\lambda)|}-\frac{l}{t}\right)\right\}\,,
\label{eq:v_and_s}
\ee
and where $\theta_{H}(x)$ is the Heaviside Theta function ($\theta_{H}(x)=0$ if $x<0$, $\theta_{H}(x)=1$ otherwise). Here, $r$ and $n$ are the indexes labeling the different species and bound states of the quasiparticles; $s_{n}^{(r)}(\lambda)$ is the Yang-Yang entropy density
\bea
s_{n}^{(r)}(\lambda)=\left(\rho_{n}^{(r)}(\lambda)+\rho_{h, n}^{(r)}(\lambda)\right) \ln \left(\rho_{n}^{(r)}(\lambda)+\rho_{h, n}^{(r)}(\lambda)\right) -\rho_{n}^{(r)}(\lambda) \ln \rho_{n}^{(r)}(\lambda)-\rho_{h, n}^{(r)}(\lambda) \ln \rho_{h, n}^{(r)}(\lambda)\,,
\eea
while $v^{(r)}_n(\lambda)$ are the quasiparticle velocities introduced in Eq.~\eqref{eq:velocities}.

We stress that, even though the above conjecture is natural, it is very important to provide numerical tests of its validity. For example, in nested systems it is known that only the first species of quasiparticles contribute to the expectation value of local conserved quantities \cite{MBPC17,PVCP18_I}; this is apparent, for instance, for the energy eigenvalues appearing in Eq.~\eqref{eq:energy}. Accordingly one could wonder whether the correct generalization of \eqref{eq:intuitive_entanglement} to the Lai-Sutherland model should only involve the first, and not both, species of the quasiparticles. 

\begin{figure}
	\centering
	\includegraphics[scale=0.22]{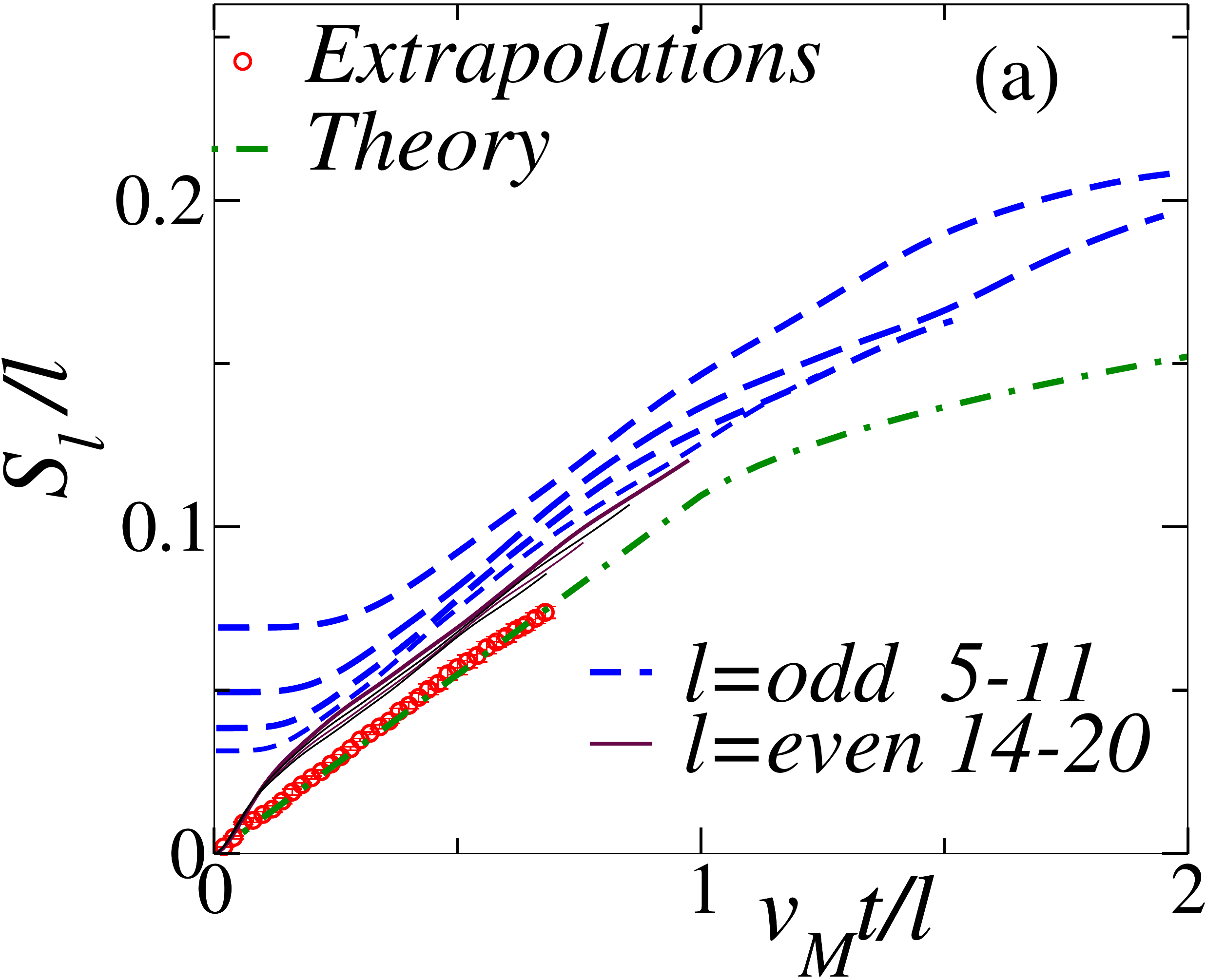} 
	\includegraphics[scale=0.22]{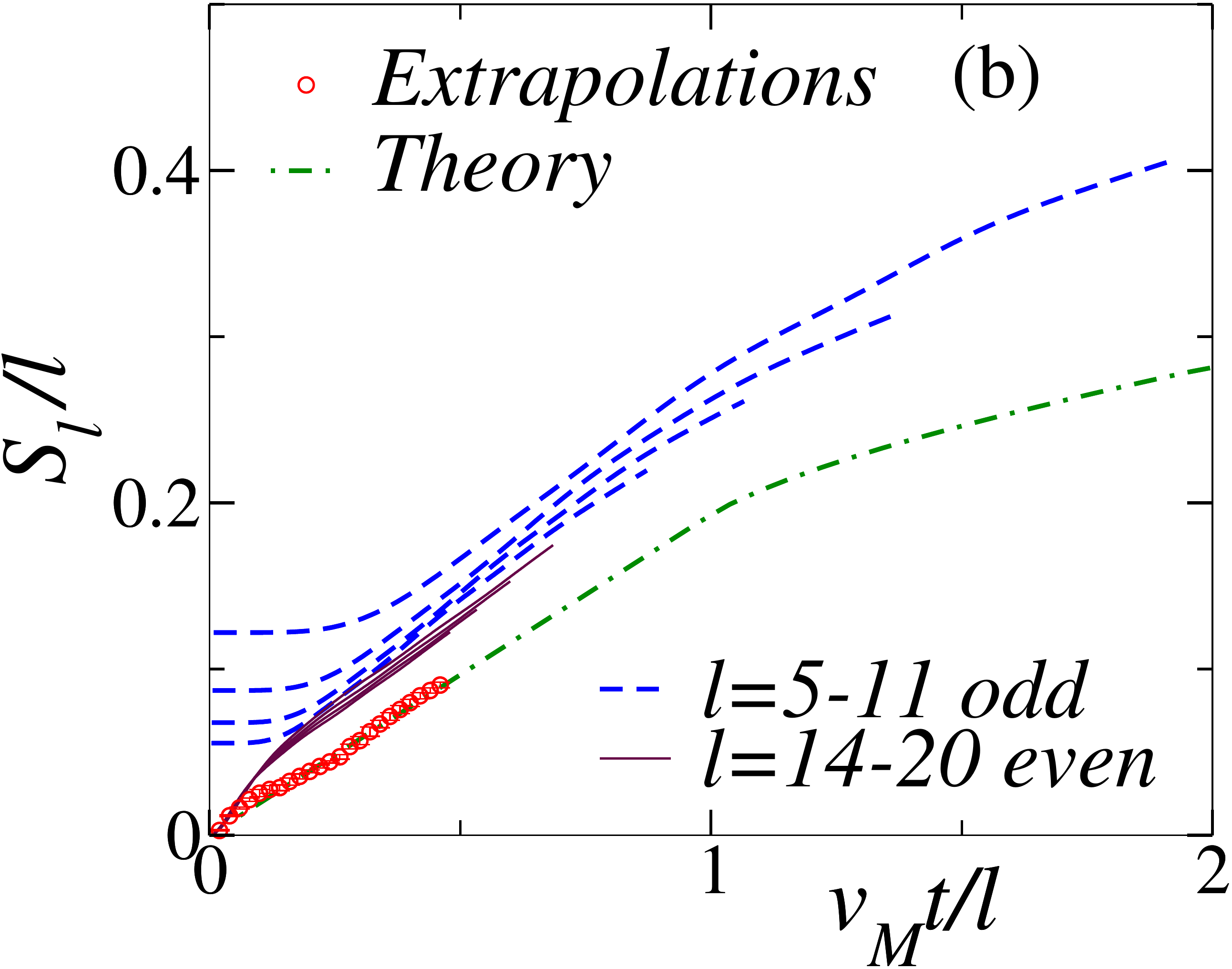} 
	\includegraphics[scale=0.22]{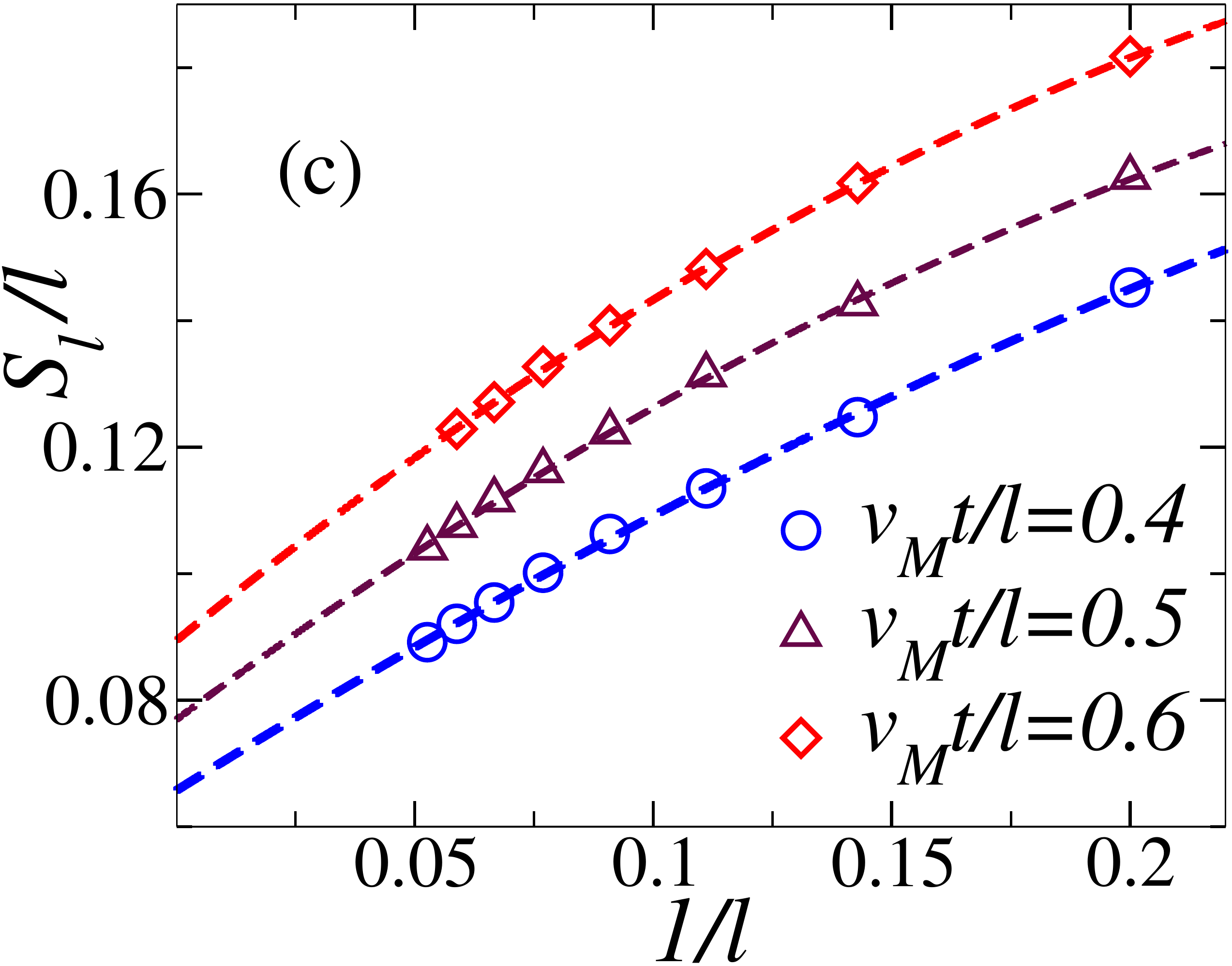} 
	\caption{Entanglement entropy after the quench from integrable initial states, in a chain of $L=48$ sites, for increases sizes $l$ of the subsystem $A$. The time axes are rescaled with $v_M/l$, where $v_M$ is the maximum velocity of the quasiparticles obtained by Bethe ansatz. Subfigure~$(a)$: the parameters of the two-site block \eqref{eq:diagonal_block} are chosen as $c_{11}=0.95$, $c_{22}=0.3$, and $c_{33}\simeq 0.087$. Dashed and solid lines correspond respectively to tDMRG results for $l=5,7,9,11$ and $l=14,16,18,20$ (from top to bottom).The dashed-dotted line is the theoretical prediction \eqref{eq:entanglemententropy}, while red circles correspond to the numerical results obtained after the extrapolation procedure. Subfigure~$(b)$: same as $(a)$, for the initial parameters $c_{11}=0.9$, $c_{22}=0.35$, and $c_{33}\simeq 0.26$. Subfigure~$(c)$:  numerical extrapolations of the tDMRG results for finite odd values of $l$ (symbols). The dashed lines correspond to the quadratic curve $S_l/l=s_{\infty}+a/l+b/l^{2}$, where $s_{\infty}$, $a$, $b$ are fitting parameters (cf. Appendix~\ref{sec:numerics}). }
	\label{fig:vonNeumann}
\end{figure}

In order to test the validity of \eqref{eq:entanglemententropy} we have computed numerically the evolution of the entanglement entropy $S_{l}(t)$ for different integrable initial states and increasing values of the subsystem size $l$. We have then performed a fit to extrapolate the value of $S_{l}(t)/l$ in the limit $t,l\to\infty$, as explained in detail in Appendix~\ref{sec:numerics}.  The numerical results are then compared against the Bethe ansatz prediction \eqref{eq:entanglemententropy}; importantly, this can be evaluated explicitly for the  integrable initial states using the analytic results of Ref.~\cite{PVCP18_I} for the corresponding post-quench quasiparticle distribution functions. We report in Fig.~\ref{fig:vonNeumann} the final outcome of this analysis, from which we see that the numerical results are in very good agreement with the analytic predictions. 
\begin{figure}
	\centering
	\includegraphics[scale=0.22]{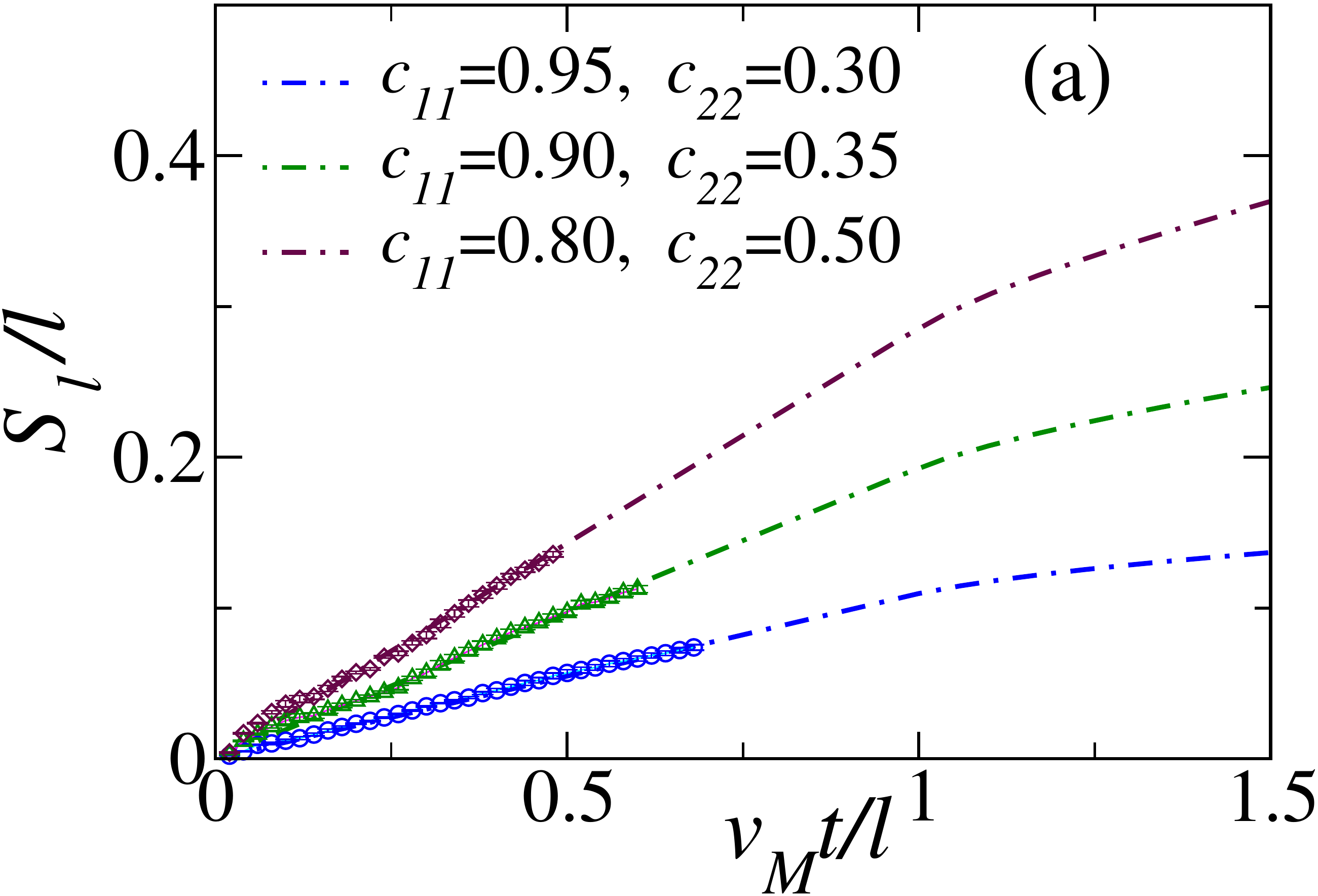} 
	\hspace{0.2cm}
	\includegraphics[scale=0.2]{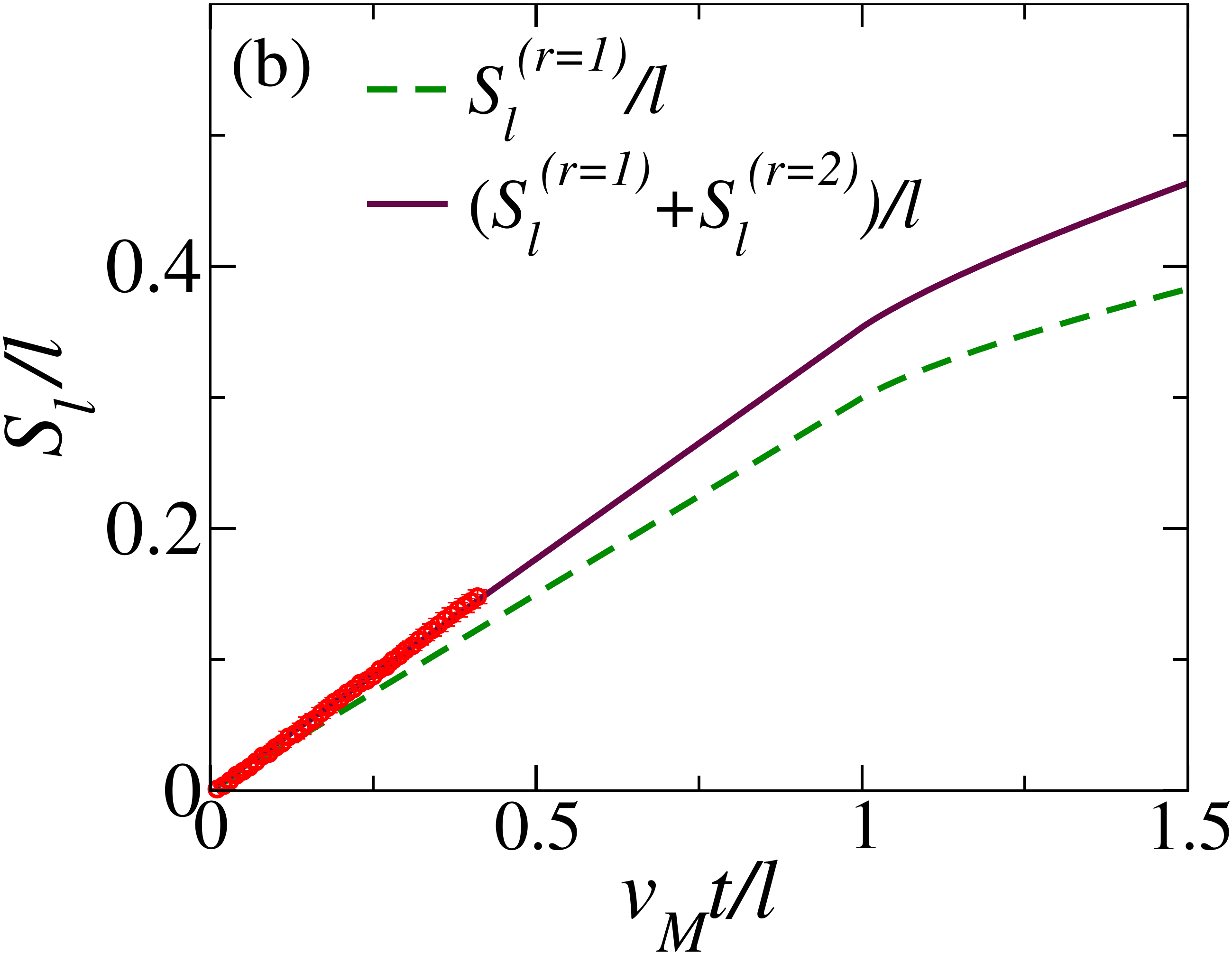} 
	\hspace{0.2cm}
	\includegraphics[scale=0.2]{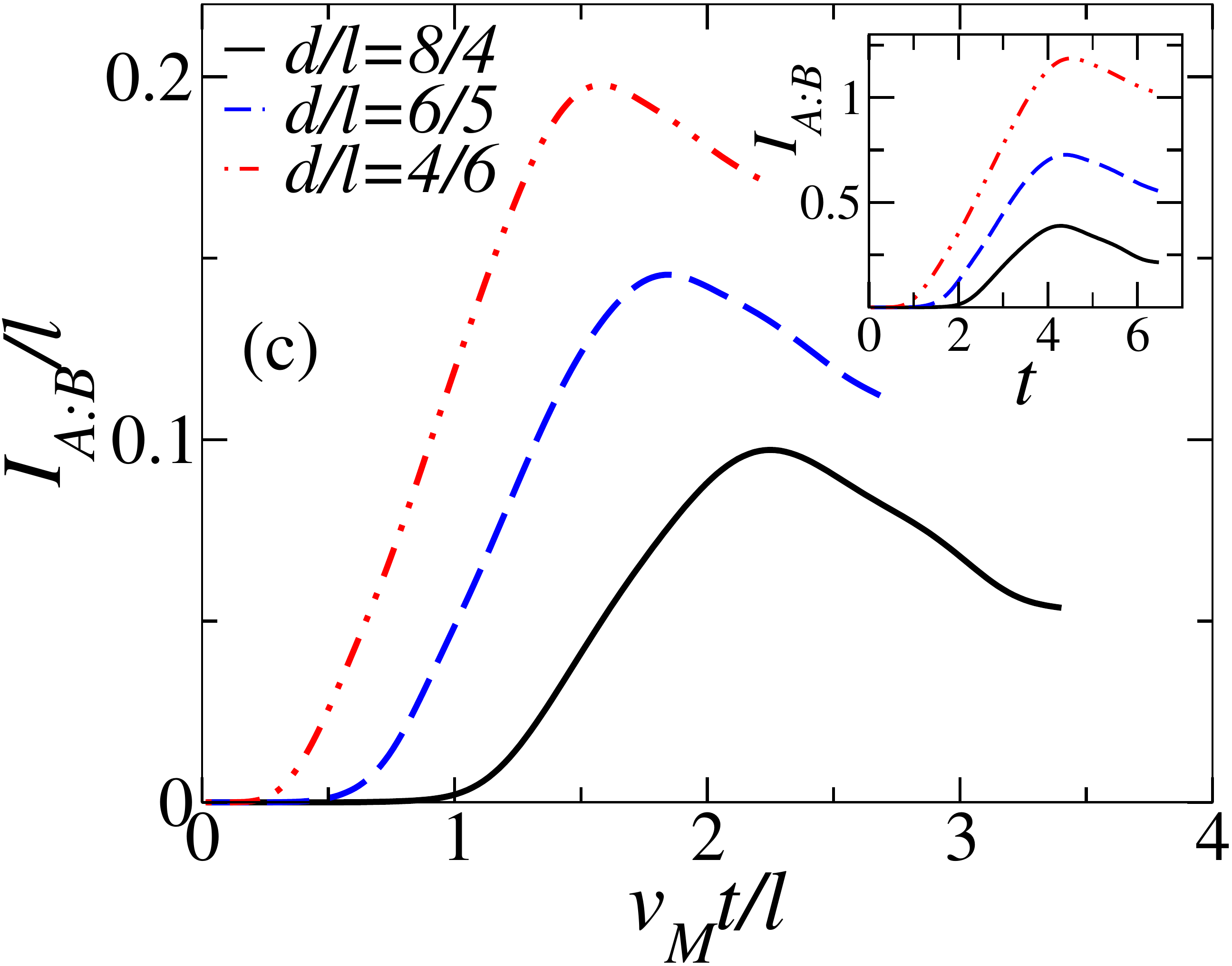} 
	\caption{Subfigure~$(a)$: extrapolated entanglement entropy after the quench from different integrable initial states (symbols). For each state, the initial parameters $c_{jj}$ of Eq.~\eqref{eq:diagonal_block} satisfy $\sum_{j=1}^3c^2_{jj}=1$. Dashed-dotted lines correspond to the theoretical prediction~\eqref{eq:entanglemententropy}. The system size is $L=48$. Subfigure~$(b)$: the extrapolated numerical data (red symbols) for the entanglement entropy are compared against the contribution $S^{(1)}_l$ of the first quasiparticle species (green dashed line). The latter is not in agreement with the numerical results; these are instead consistent with the prediction \eqref{eq:entanglemententropy} (solid line), which takes into account both species of quasiparticles. The  initial parameters chosen for this plot are $c_{11}=c_{22}=0.6$, $c_{33}\simeq 0.53$, while the system size is $L=48$. Subfigure $(c)$: rescaled mutual information $I_{A:B}/l$ against $v_M t/l$ for the initial state corresponding to $c_{11} = 0.95$, $c_{22 }= 0.3$, $c_{33}\simeq 0.087$. Inset: the same data are displayed without rescaling the axes. The system size is $L=16$, while the curves correspond (from bottom to top) to $l=4$, $5$, $6$.}
	\label{fig:extrapolated}
\end{figure}

Although the fitting procedure to extrapolate the infinite-time limit bears a numerical error, the method is accurate enough to clearly resolve the entanglement evolution for different initial states, as shown in Fig.~\ref{fig:extrapolated}$(a)$. We have also tested our numerics against the contribution $S_l^{(1)}(t)$ of the first quasiparticle species in Eq.~\eqref{eq:entanglemententropy}: this is displayed in Fig.~\ref{fig:extrapolated}$(b)$, where we see that the numerical data are in agreement with the prediction \eqref{eq:entanglemententropy}, which takes into account the contribution of both species, but not with the individual term $S_l^{(1)}$. This rules out the possibility of a picture where only the first species of quasiparticles contributes to the entanglement dynamics. Overall, our analysis fully corroborates the validity of Eq.~\eqref{eq:entanglemententropy} as the correct generalization of the results of Ref.~\cite{AlCa17,AlCa18} to the case of nested spin chains.

The knowledge of the post-quench rapidity distribution functions corresponding to the integrable initial states also allows us to provide predictions for other quantities related to the entanglement entropy, such as the mutual information between 
two subsystems $A$ and $B$, which has been recently proposed as a simple measure of scrambling~\cite{asplund,AlCa19}. We recall that the mutual information is defined as $I_{A:B}=S_{A}+S_{B}-S_{A\cup B}$, where $S_{A}$, $S_{B}$, and $S_{A\cup B}$ are the von Neumann entanglement entropies of the subsystems $A$, $B$ and of their union, $A\cup B$, respectively. By employing the very same quasiparticle picture discussed above, one can derive the following formula~\cite{MBPC17}
\begin{equation}
\begin{aligned} 
I_{A : B}(t)=& \sum_{r=1,2} \sum_{n=1}^{\infty} \int \mathrm{d} \lambda\left[\left(2\left|v_{n}^{(r)}(\lambda)\right| t-d\right) \chi_{[d, d+l]}\left(2\left|v_{n}^{(r)}(\lambda)\right| t\right)\right.\\ &+\left(d+2 l-2\left|v_{n}^{(r)}(\lambda)\right| t\right) \chi_{[d+l, d+2 l]}\left(2\left|v_{n}^{(r)}(\lambda)\right| t\right) ] s_{n}^{(r)}(\lambda)\,,
\end{aligned}
\label{eq:mutual_info}
\end{equation}
where $\chi_{[a,b]}(x)$ is the characteristic function of $[a, b]$, i.e. it is equal to $1$ if $x\in [a, b]$ and equal to $0$ otherwise. Here we assumed $A$ and $B$ to be of the same size $l$, and we denoted their distance by $d$, while $s^{(r)}_{n}(\lambda)$ and $v^{(r)}_{n}(\lambda)$ are defined as in Eq.~\eqref{eq:v_and_s}. We stress that Eq.~\eqref{eq:mutual_info} is conjectured to be exact in the limit $l,d,t\to\infty$, while keeping the ratios $l/t$ and $d/t$ constant. The mutual information is interesting as its time dependence signals, at large times, the presence of different species and bound states of quasiparticles, manifesting themselves as subsequent peaks of $I_{A:B}(t)$~\cite{MBPC17,PVCP18_I}. Unfortunately, the entanglement entropy of multiple disjoint intervals in tDMRG simulations is computationally more demanding, when compared to the entropy of a single interval. Hence, we are able to provide reliable numerical data only for small system and subsystem sizes. An example is shown in Fig.~\ref{fig:extrapolated}$(c)$, where $I_{A:B}(t)/l$ is plotted as a function of $v_{M}t/l$ for a given initial integrable state (here $v_{M}$ is the maximum velocity of the quasiparticles, as computed via Bethe ansatz).  From the figure we see that $I_{A:B}(t)\simeq 0$ up to a time  $t\simeq d/2v_{M}$, after which it increases linearly, as expected from Eq.~\eqref{eq:mutual_info}; however, the time scales and system sizes that we can reach are too small to be compared with the theoretical prediction~\eqref{eq:mutual_info}, and to detect the peaks associated with the different species and the bound states of the quasiparticles~\cite{MBPC17, PVCP18_I}. 

Finally, with the same numerical procedure we have also computed the evolution of the R\'enyi entropies~\eqref{eq:Renyie} for different values of $\alpha$. Since there is not yet a Bethe ansatz prediction for their post-quench dynamics \cite{AlCal-b}, the corresponding numerical results are reported in Appendix~\ref{sec:numerics}, to which we refer the interested reader.

\section{Evolution of correlation functions}
\label{sec:correlations}
\begin{figure}
	\centering
	\includegraphics[scale=0.23]{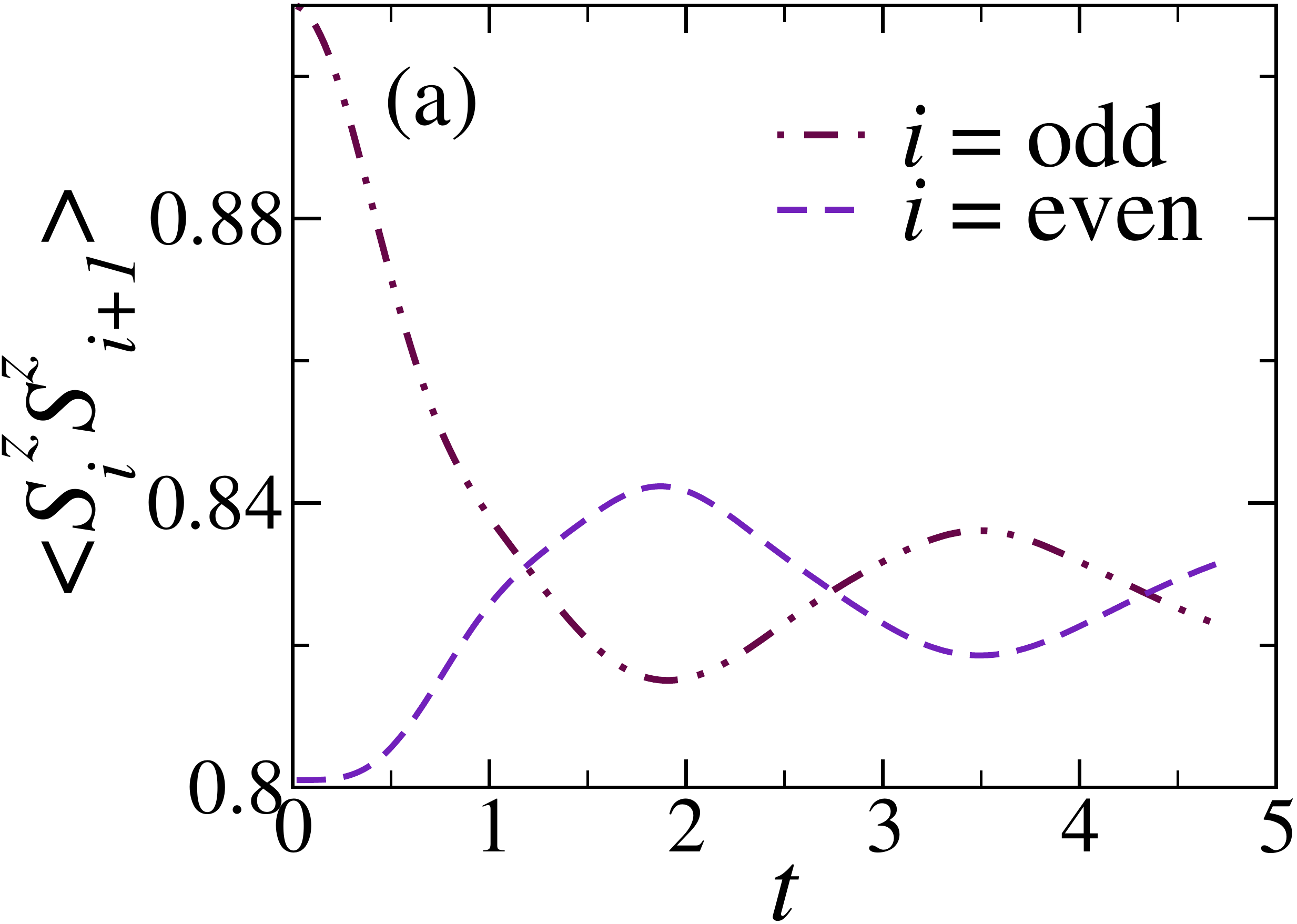} 
	\includegraphics[scale=0.23]{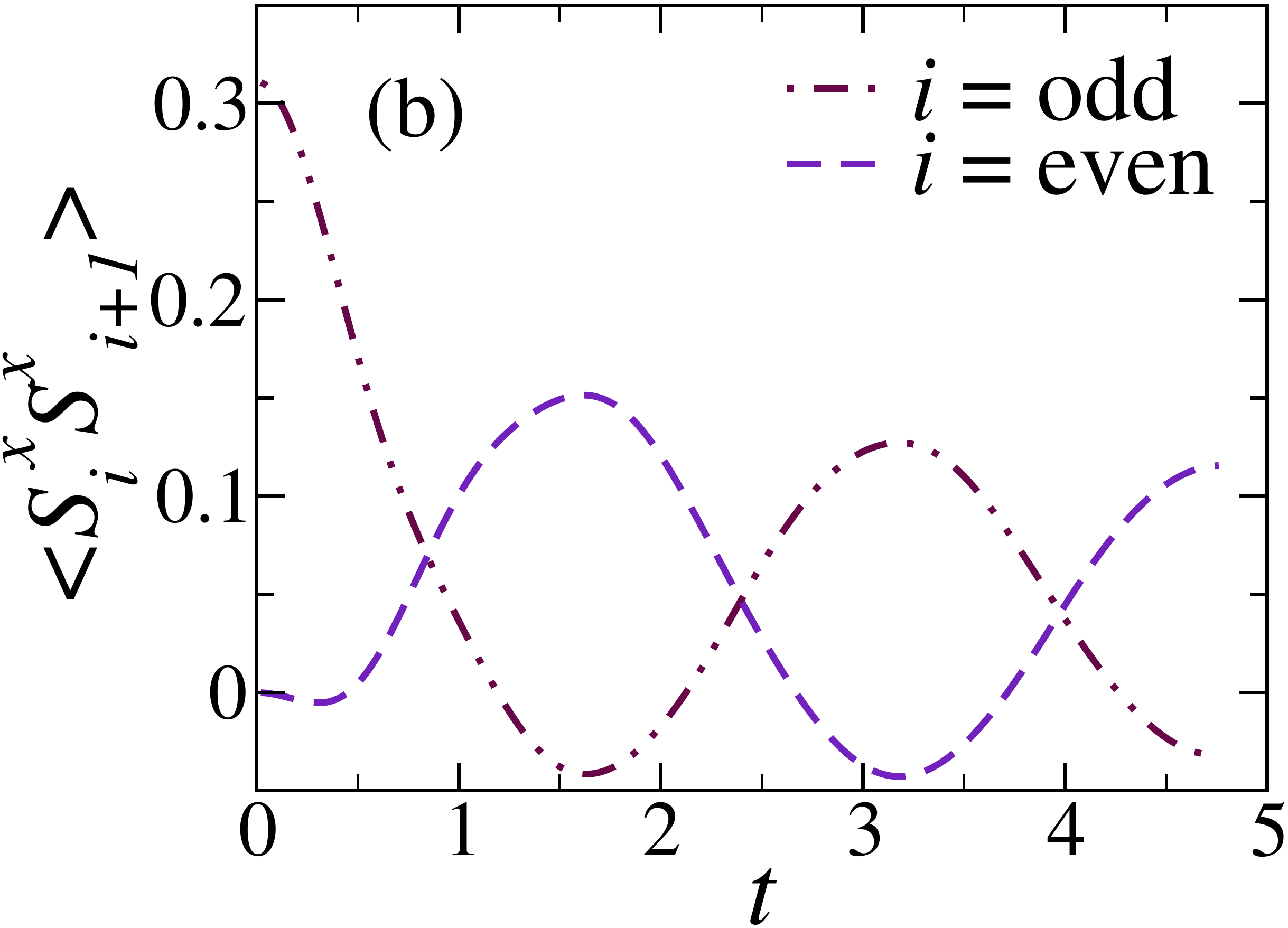} 
	\includegraphics[scale=0.23]{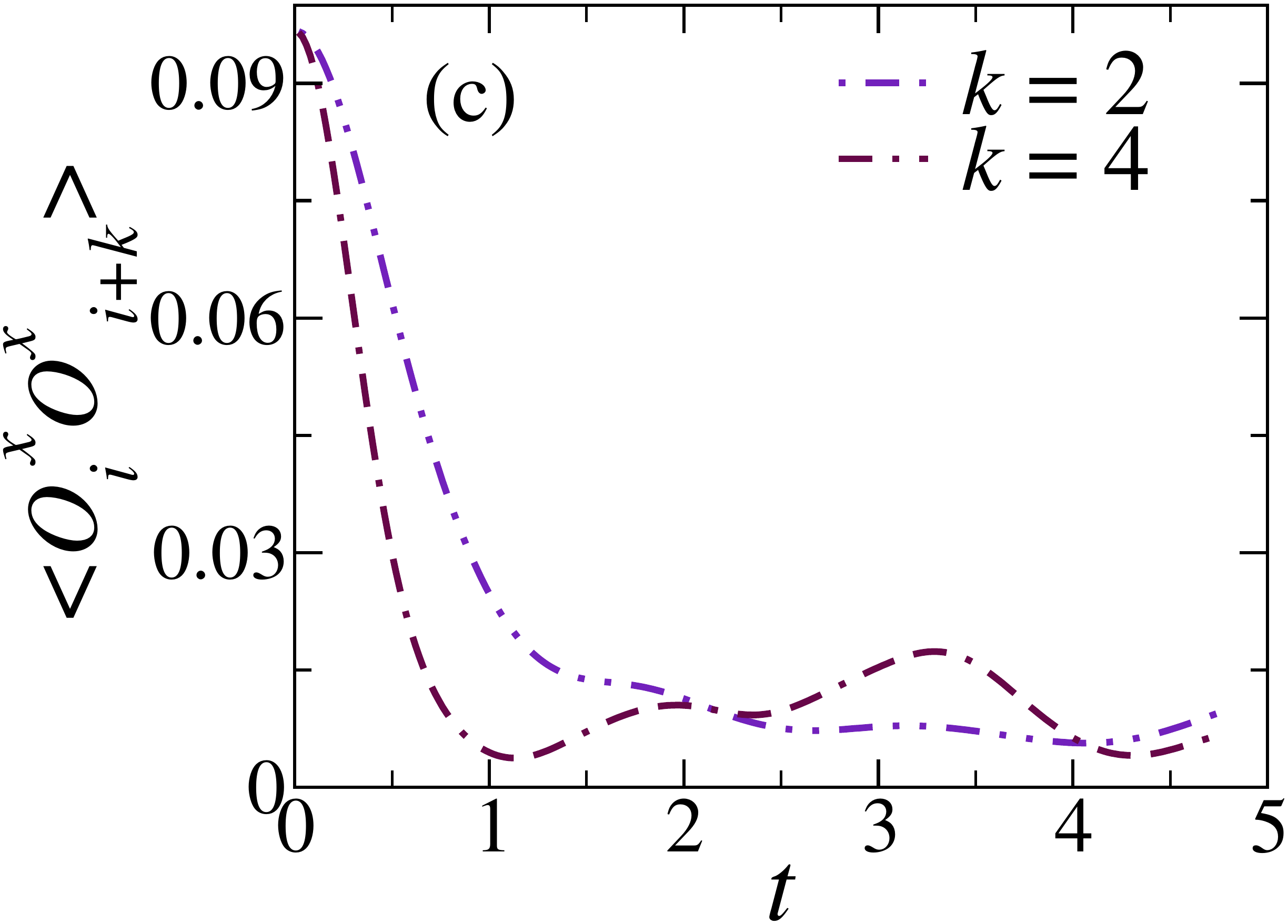}
	\caption{Subfigures $(a)$ and $(b)$: dynamics of local correlation functions $\braket{S^{z}_iS^{z}_{i+1}}$ and $\braket{S^{x}_iS^{x}_{i+1}}$ after an integrable quench, with initial parameters $c_{11}=0.95$, $c_{22}=0.3$, and $c_{33}\simeq 0.087$. The tDMRG simulations have been carried out for a chain of $L=60$ sites, while we chose $i=L/2-1$ for $i$ odd, and  $i=L/2$ for $i$ even. Subfigure $(c)$: evolution of the correlation function $\braket{O^x_iO^x_{i+k}}$, where $O^x_i=S^x_iS^x_{i+1}$, for the same quench.}
	\label{fig:local_correlators}
\end{figure}

In this section, we study another important aspect of the post-quench dynamics, namely the evolution of local correlation functions. It is useful to start by observing a few properties of the Hamiltonian \eqref{eq:hamiltonian} and of the integrable initial states \eqref{eq:initial_state}. Focusing on the site $k=L/2$, we notice that the Hamiltonian is invariant under the ``bond-inversion symmetry'' $P_B$ around the ``link'' between $k$ and $k+1$, acting on local operators as $O_{j}\to O_{L+1-j}$, where $O_j$ acts as the identity on $h_n$ for $n\neq j$. The same transformation leaves invariant an integrable state $\ket{\Psi_0}$ [with two-site block \eqref{eq:diagonal_block}], namely $P_B\ket{\Psi_0}=\ket{\Psi_0}$. Finally, using that $P_BO_{L/2}P_B^{-1}=O_{L/2+1}$, this implies
\be
\braket{\Psi_0|O_{L/2}(t)|\Psi_0}=\braket{\Psi_0|O_{L/2+1}(t)|\Psi_0}\,.
\ee
In the case of periodic boundary conditions, one could repeat the same argument for each site in the chain, obtaining
\be
\braket{\Psi_0|O_{j}(t)|\Psi_0}=\frac{1}{L}\sum_{k=1}^{L}\braket{\Psi_0|O_{k}(t)|\Psi_0}\,.
\label{eq:translational_invariance}
\ee
We stress that \eqref{eq:translational_invariance} holds despite $\ket{\Psi_0}$ is not translationally invariant. As a result, due to the $SU(3)$ symmetry of the Hamiltonian \eqref{eq:hamiltonian}, the expectation value of point-wise operators $O_j$ is conserved after the quench from integrable initial states, as it is easily shown using \eqref{eq:translational_invariance}.

\begin{figure}
	\centering
	
	\includegraphics[height=1.65in,width=2.2in]{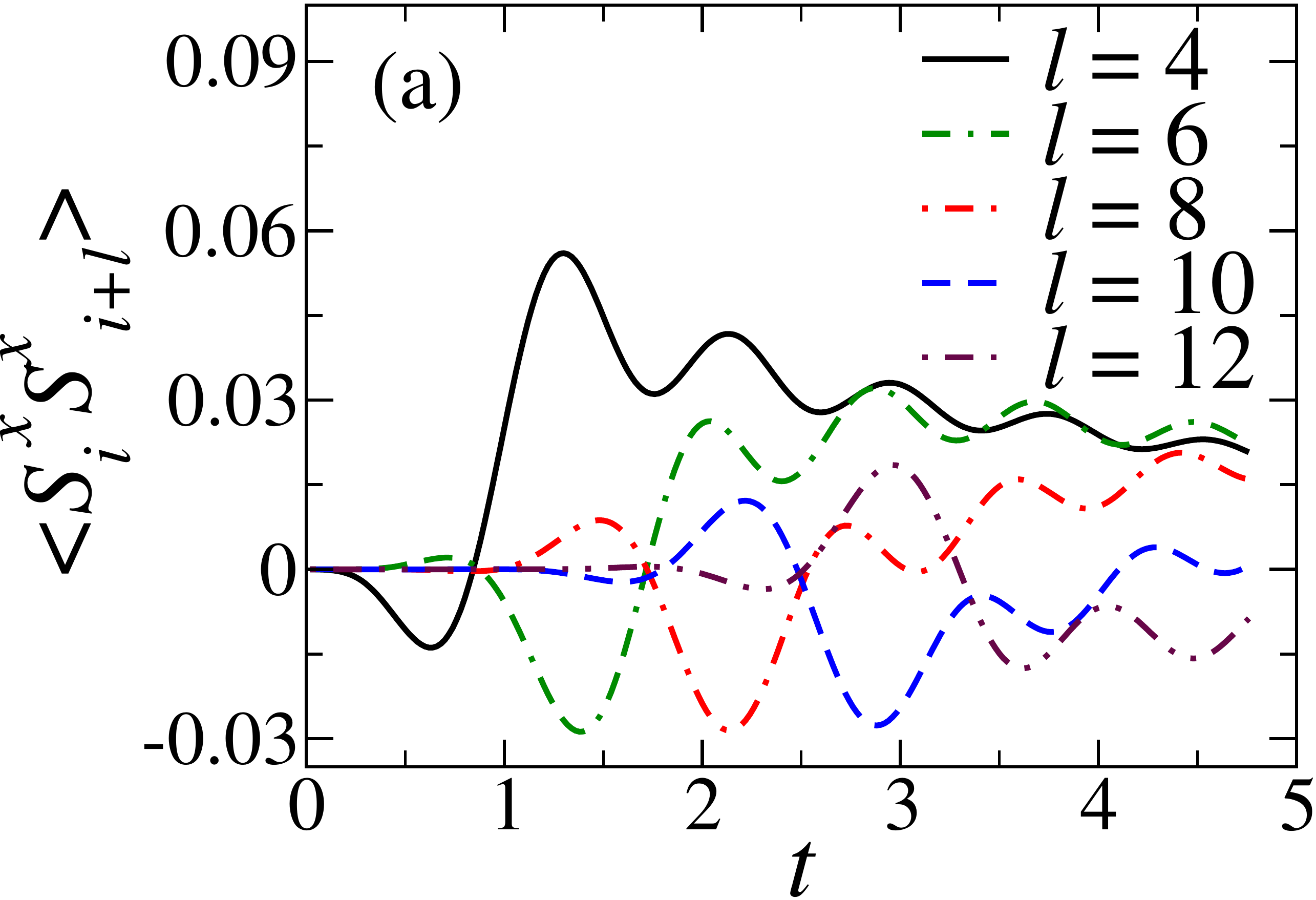} 
	\includegraphics[height=1.65in,width=2.2in]{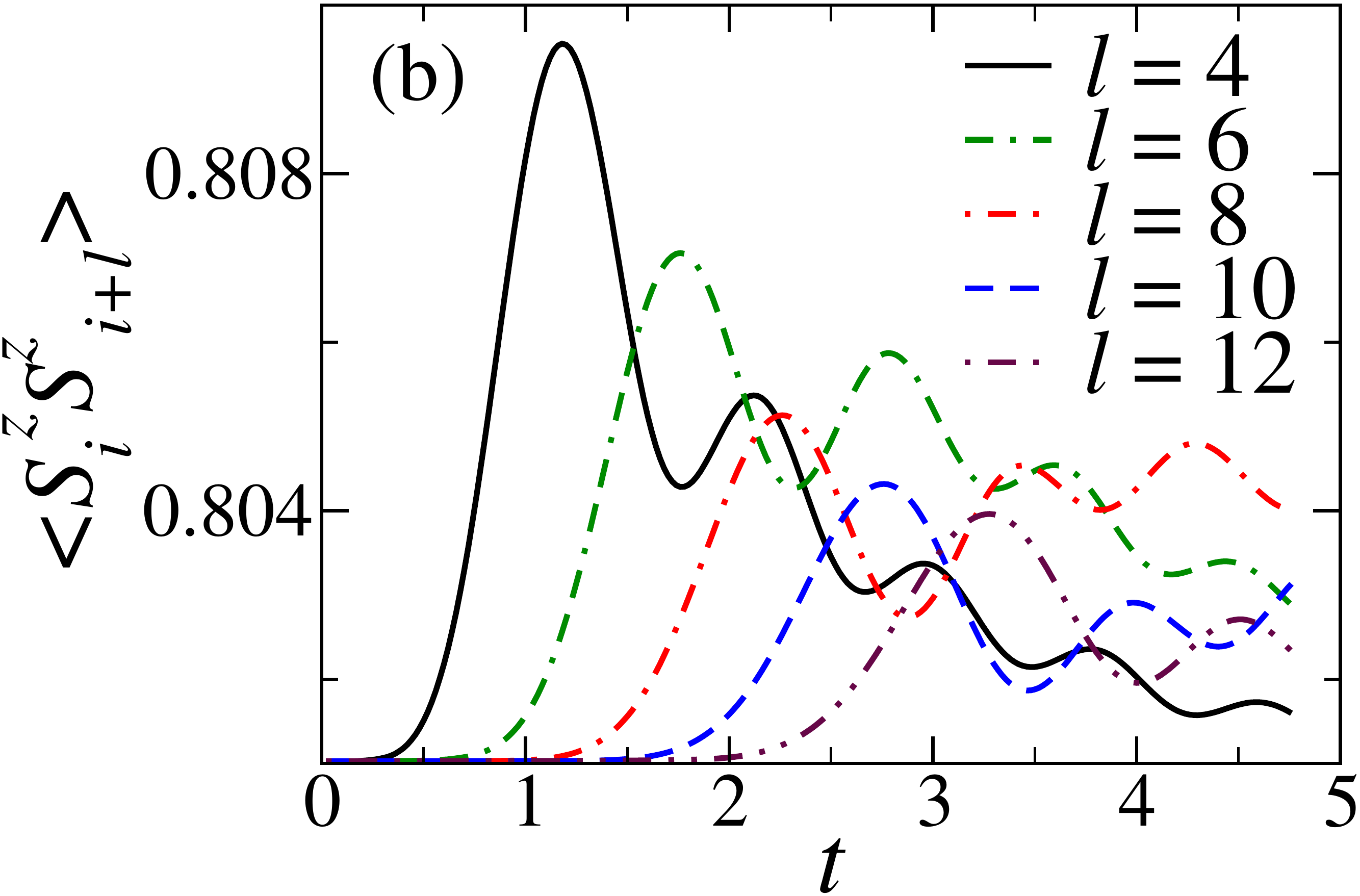} 
	\includegraphics[height=1.75in,width=2.5in]{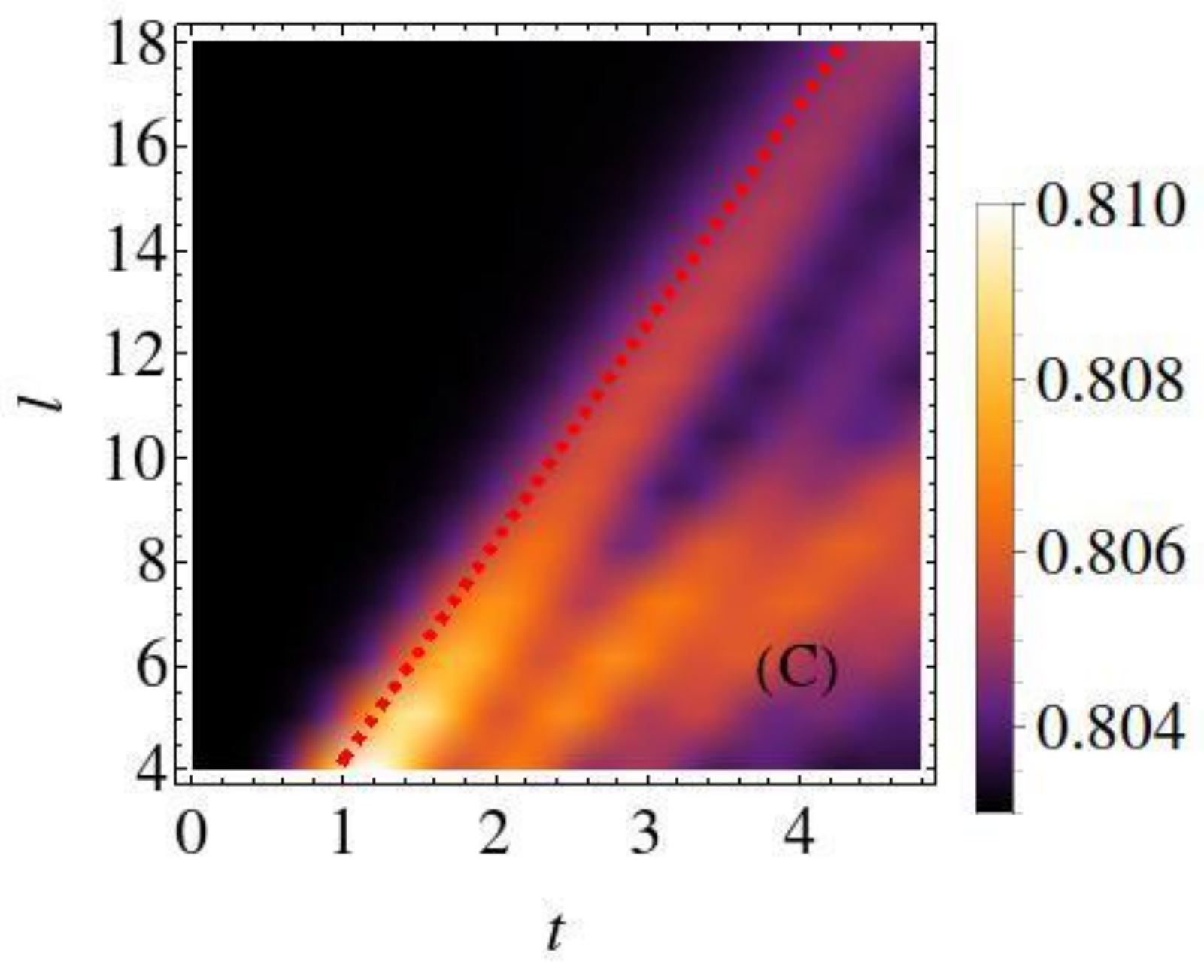} 
	\caption{Subfigures $(a)$ and $(b)$:  correlation functions $\braket{S^{x}_iS^{x}_{i+l}}$ and $\braket{S^{z}_iS^{z}_{i+l}}$ for increasing $l$, after a quench from the integrable state corresponding to $c_{11}=0.95$, $c_{22}=0.3$, and $c_{33}\simeq 0.087$. The tDMRG simulations have been carried out for a chain of $L=60$ sites, while we chose $i=L/2$. Subfigure $(c)$: space-time plot for transverse spin correlation function $\braket{S^{z}_iS^{z}_{i+l}}$ and the same initial state, exhibiting a clear light-cone structure. The dashed red line corresponds to the maximum velocity $v_M$ of the quasiparticles, obtained  from Bethe ansatz (for this state $v_M\simeq 2.10$).}
	\label{fig:light_cone_examples}
\end{figure}

According to the discussion above, the simplest non-trivial correlation functions that we can consider are two-site correlators.  We start by studying $\braket{S^{z}_j(t)S^{z}_{j+1}(t)}$ and $\braket{S^{x}_j(t)S^{x}_{j+1}(t)}$, which we have computed numerically for a quench from different initial states; an example is reported in Fig.~\ref{fig:local_correlators}. In each case, the correlators depend on the parity of $j$, due to the two-site structure of the initial state (which breaks translational invariance). From Fig.~\ref{fig:local_correlators} we see that within the available time scales the two-site correlation functions exhibit large oscillations; furthermore, these appear to be around the same value for even and odd $j$, suggesting the expected restoration of translational symmetry at large times~\cite{FCEC14}. We have found that oscillations appear to be suppressed in the evolution of four-point functions, as shown in Fig.~\ref{fig:local_correlators}$(c)$. In any case, our numerical data do not allow us to extract the asymptotic value of the local correlation functions. Furthermore, while exact result have been recently derived in the thermal case to compute expectation values of local observables in the Lai-Sutherland model~\cite{RiKl19}, such results have not yet been generalized to the case of GGE-states, as in the case of $XXZ$ Heisenberg chains \cite{MePo14,Pozs17}. For these reasons, it appears problematic to test the validity of the analytic results of Ref.~\cite{PVCP18_I,PVCP18_II} for the post-quench rapidity distribution functions, based on the asymptotics of local correlations.

Luckily, one can perform another test, albeit more indirect, which can be carried out within the time scales accessible to our numerical scheme. This is based on the computation of the ``light-cone'' velocity of correlations~\cite{BoEL14, FSEH13}, namely the speed with which correlations spread after a quench. First, it is useful to recall that the Bethe ansatz provides us with a prediction for the latter in the scaling limit of large distances and times, based on the knowledge of the post-quench rapidity distribution functions. Let us consider $\braket{O_j(t)O_{j+l}(t)}$, where $O_x$ is an observable localized at site $x$. According to the quasiparticle picture, correlations between $O_j$ and $O_{j+l}$ are caused by pairs of quasiparticles produced in the same spatial point and that arrive at positions $j$ and $j+l$ at the same time. From this picture, we obtain that the time needed for these local operators to be correlated is at least $t=l/(2v_M)$, where $v_M$ is the maximum velocity of the quasiparticles. This can be computed by Eq.~\eqref{eq:velocities}, once the functions $\rho^{(r)}_n(\lambda)$ are known, so that the numerical computation of time-evolved correlators provides a test for the validity of the results of Refs.~\cite{PVCP18_I,PVCP18_II} for the post-quench rapidity distribution functions. 

\begin{figure}
	\centering
		\includegraphics[scale=0.2]{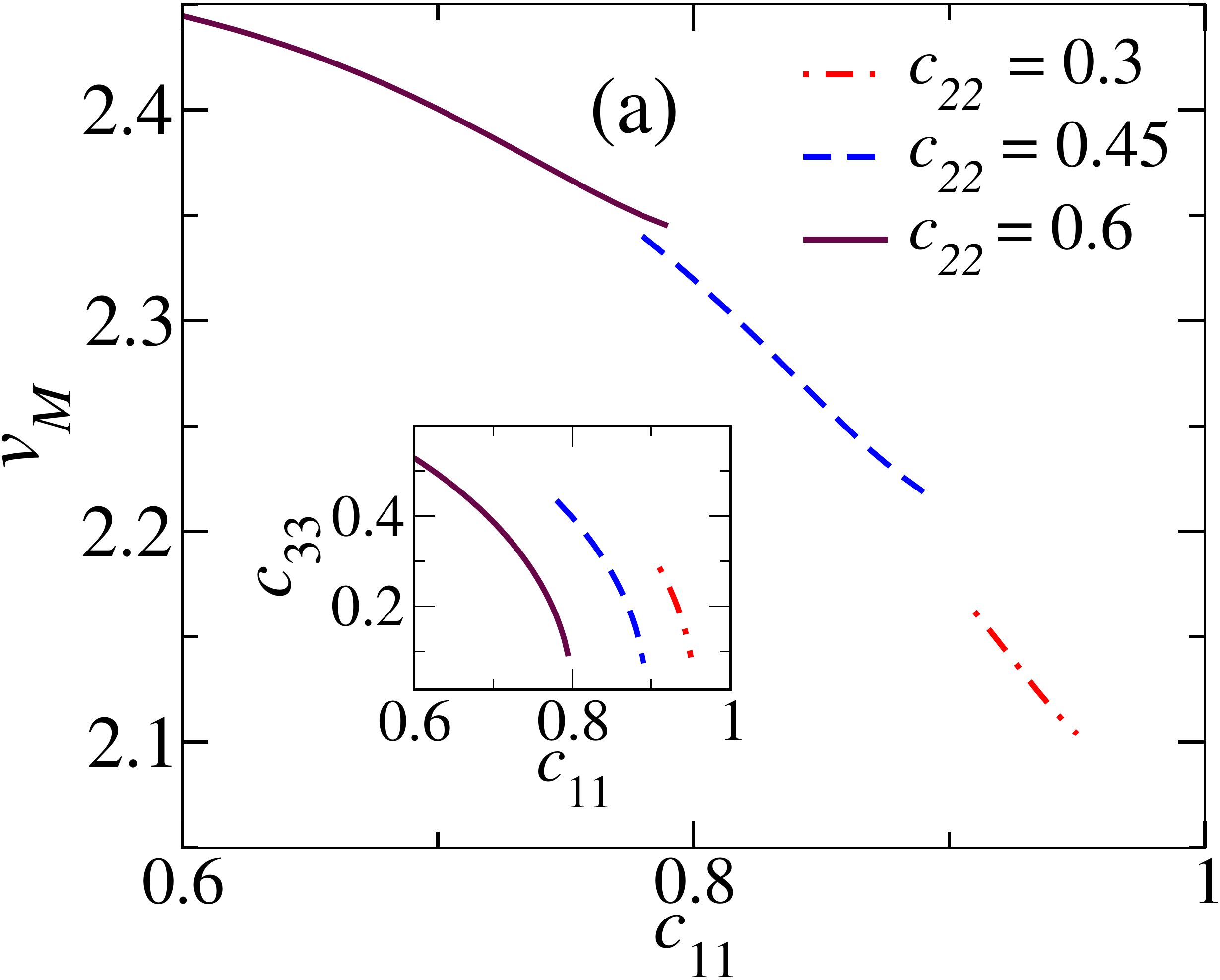} 
	\includegraphics[scale=0.22]{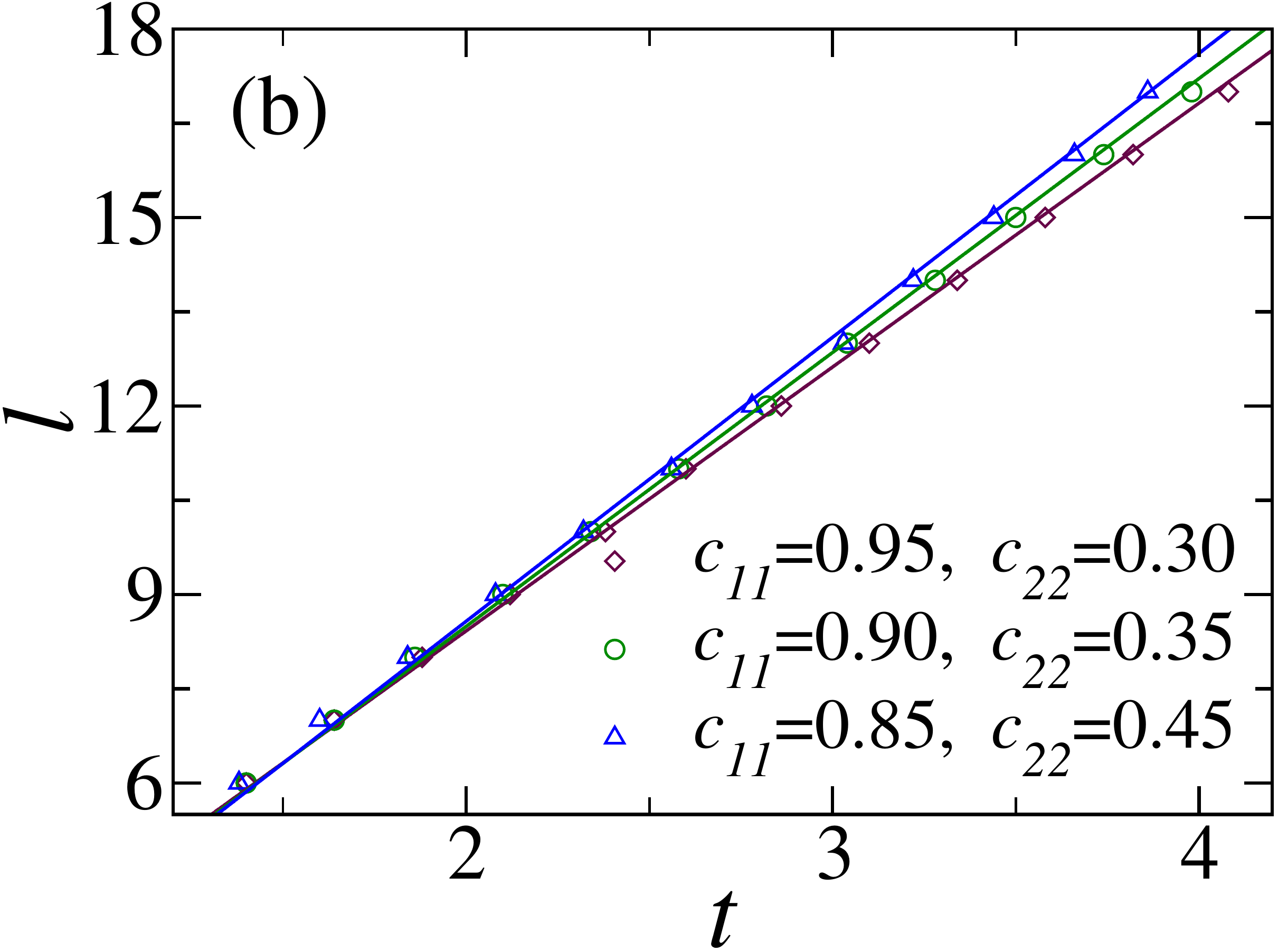} 
		\includegraphics[scale=0.22]{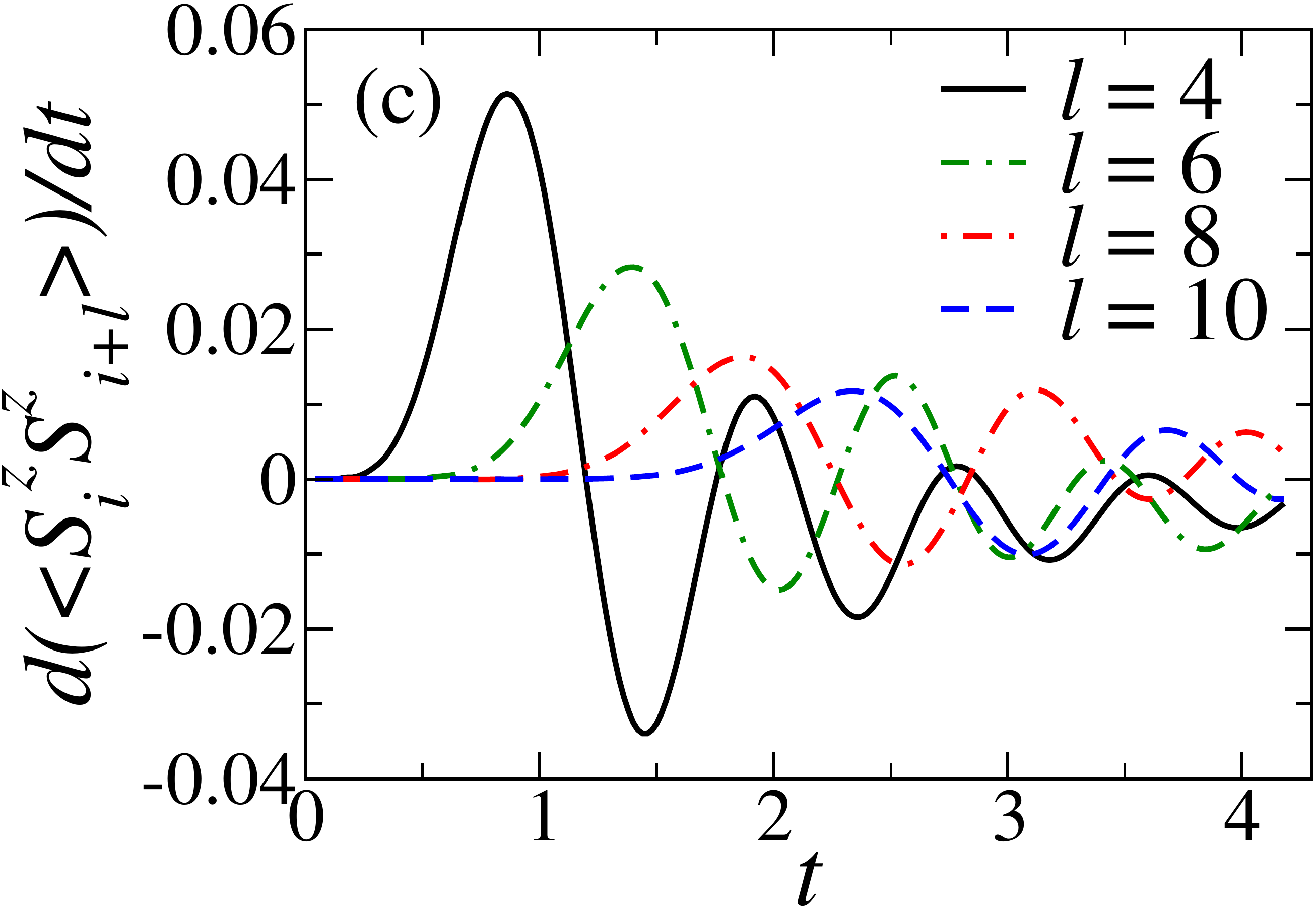} 
	\caption{Subfigure~$(a)$: maximum velocity of the quasiparticles, for different initial states parameterized by the two-site block \eqref{eq:diagonal_block}. For each state, we choose the normalization $c^2_{11}+c^2_{22}+c^2_{33}=1$. The three lines correspond to different choices of $c_{22}$, while the inset shows the variation of $c_{33}$, as $c_{11}$ increases. Subfigure~$(b)$: numerical results for the first inflection points of $\braket{S^z_jS^z_{j+l}}$ for increasing values of $l$ (symbols). Solid lines correspond to the curves $2v_Mt+c$, where $c$ are a fitting constants. Subfigure $(c)$: time derivative of $\braket{S^z_jS^z_{j+l}}$ for increasing $l$, after the quench from the integrable state with parameters $c_{11}=0.9$, $c_{22}=0.35$ and $c_{33}\simeq 0.26$. The set of the first maxima (i.e. the first inflection points of $\braket{S^z_jS^z_{j+l}}$) are identified as the arrival times of the light-cone. 
	}
	\label{fig:velocity}
\end{figure}

In order to probe the light-cone velocity of the correlation functions, we compute $\braket{S^{\alpha}_j(t)S^{\alpha}_{j+l}(t)}$ for $\alpha=x,z$ and increasing values of $l$. Examples of our results for a particular integrable initial state are reported in the plots of Fig.~\ref{fig:light_cone_examples}, from which one can immediately see the emergence of a light-cone structure.  Quantitative results for the corresponding velocity can be obtained from these plots following the procedure employed in Ref.~\cite{BoEL14, MWNM09}: in particular, for a given correlator $\braket{S^{\alpha}_j(t)S^{\alpha}_{j+l}(t)}$ one can identify the first ``inflection point'' [i. e. the first maximum of the its time-derivative, cf. Fig~\ref{fig:velocity}$(c)$] as the arrival time $t_l$ of the light-cone and plot the sequence $t_l$ as a function of $l$. Then, the velocity of the light-cone can be extracted as the slope of the straight line fitting the set of points $(t_l,l)$ (which should be independent of the particular correlation function which has been chosen, as we have explicitly verified).

A summary of our results is displayed in Fig.~\ref{fig:velocity}. First, in subfigure $(a)$ we report the Bethe ansatz predictions for the maximum quasiparticle velocity corresponding to different initial states. We see that the latter has a weak dependence on the initial parameters $c_{jj}$. Furthermore, states with large velocity have large entanglement and are more difficult to simulate. Hence, the expected maximum velocity of the states that we can study numerically vary in a rather restricted range. In Fig.~\ref{fig:velocity} we display our results for the light-cone velocity for three different initial states, from which the aforementioned weak dependence on the state parameters is apparent. The numerical data are compared against the ansatz $y(t)=2v_Mt+c$, where we have performed a fit to determine the constant $c$. We see that they are in good quantitative agreement with the analytic predictions, and that these are able to resolve the different initial states, despite the weak dependence on the quench parameters. In conclusions, our results are always seen to corroborate the validity of the Bethe ansatz predictions for the light-cone velocity of local correlations.

\section{Conclusions}
\label{sec:conclusions}

In this work we have presented numerical results for quantum quenches in the nested Lai-Sutherland model from a family of integrable initial states, focusing on the spreading of both correlation functions and entanglement entropy. By means of tDMRG simulations, we tested explicitly the validity of the nested version of the conjectured quasiparticle formula for the growth of the entanglement entropy of Refs.~\cite{AlCa17,AlCa18}, and the Bethe ansatz predictions for the ``light-cone'' velocity of correlation functions~\cite{BoEL14}. The present paper complements other studies in the literature, extending to the case of nested models the body of numerical evidence corroborating the existing theory of integrable systems out of equilibrium~\cite{CaEM16}.

The theoretical predictions tested in this work are based on the analytic results of Ref.~\cite{PVCP18_I,PVCP18_II}, where the post-quench rapidity distribution functions for the integrable states were computed. While the small initial entanglement of these states allowed us to simulate their dynamics by means of the tDMRG algorithm, we have seen that the correlation functions exhibit large relaxation times, so that it is not possible to extract their asymptotic behavior based on short and intermediate-time simulations. In this respect, it would be highly desirable to generalize the string-charge duality method~\cite{IQDB16} to nested systems: the latter would allow us to enlarge the class of initial states for which analytic predictions can be made, hopefully including states with faster local relaxation. Another open problem with strong connections to our work is the derivation of analytic formulas to compute GGE correlation functions in nested models, analogously to what has been done in the case of Heisenberg chains~\cite{MePo14}. These would make it possible to give predictions for the asymptotics of local observables based on the knowledge of the post-quench rapidity distribution functions and allow for a direct comparison with numerical results. We hope that our work will serve as a motivation for future investigations in these directions.

\section*{Acknowledgments}

L.~P. acknowledges support from the Alexander von Humboldt foundation and from the Deutsche Forschungsgemeinschaft under Germany's Excellence Strategy – EXC-2111 – 390814868. P.~C. acknowledges support from ERC under Consolidator grant  number 771536 (NEMO).
P. C. and L. P. acknowledge Eric Vernier and Bal\'asz Pozsgay for discussions and collaboration on related subjects. 

\appendix

\section{Details on the numerical computations}
\label{sec:numerics}

In this appendix we provide details on the numerical methods that we have employed for the computation of the von Neumann and R\'enyi entanglement entropies, and present additional plots for the latter.

All the data displayed in the main text are obtained using the tDMRG algorithm, as implemented in the iTensor library~\cite{itensor}, where open boundary conditions have always been chosen. The data in Figs.~\ref{fig:vonNeumann} and \ref{fig:extrapolated} are obtained by considering a chain of $L=48$ sites, and the subsystem is always chosen at one of its  edges. Due to open boundary conditions, the numerical data are compared with the Bethe ansatz formula in Eq.~\eqref{eq:entanglemententropy} with the substitution $t\to t/2$, which corresponds to the fact that the quasiparticles can only enter from one of the two borders of the subsystem under consideration~\cite{AlCa17}. We choose the maximum bond dimension $\xi=1500$ and a time step $\Delta t=0.02$ for all simulations. We have checked the robustness of our results by performing simulations for different values of $\Delta t$, $\xi$ and 
$L$. All the entanglement entropy data reported in the manuscript are for $L=48$, which is the maximum system size we could reach with our computational resources. Furthermore, all of our results have been checked against exact diagonalization calculations for small times.

For the initial states in Fig.~\ref{fig:vonNeumann}, the entanglement entropy $S_l(t=0)$ is different, depending on whether the subsystem considered contains an odd or an even number of sites. More precisely, it is a non-zero constant for $l$ odd, while it is vanishing for $l$ even. Of course, in both cases one has $\lim_{l\to \infty }S_l(t=0)/l=0$. In order to extrapolate the value of $S_l/l$ to the thermodynamic limit, we have performed a fit of the data (at fixed $v_Mt/l$) using the function 
\begin{equation}
	\frac{S_l}{l}=s_{\infty}+\frac{a}{l}+\frac{b}{l^{2}}\,,
\end{equation}
where $s_{\infty}$, $a$ and $b$ are fitting parameters. Given the even/odd effect in the finite-size data, we have performed the fit independently for the even and odd sequences, obtaining different numerical results $s^{(e/o)}_{\infty}$. The (small) differences between the fits for even and odd values of $l$ are attributed to finite size effects; hence, as the best estimate for $s_{\infty}$ we chose the average $(s^{(e)}_\infty+s^{(o)}_\infty)/2$. We note that the numerical difference between $s^{(e)}_\infty$ and $s^{(o)}_\infty$ was found to decrease as $t/l$ increased for our set of finite-size data, consistently with the expected restoration of translational symmetry at large times.

\begin{figure}
	\centering
	\includegraphics[scale=0.22]{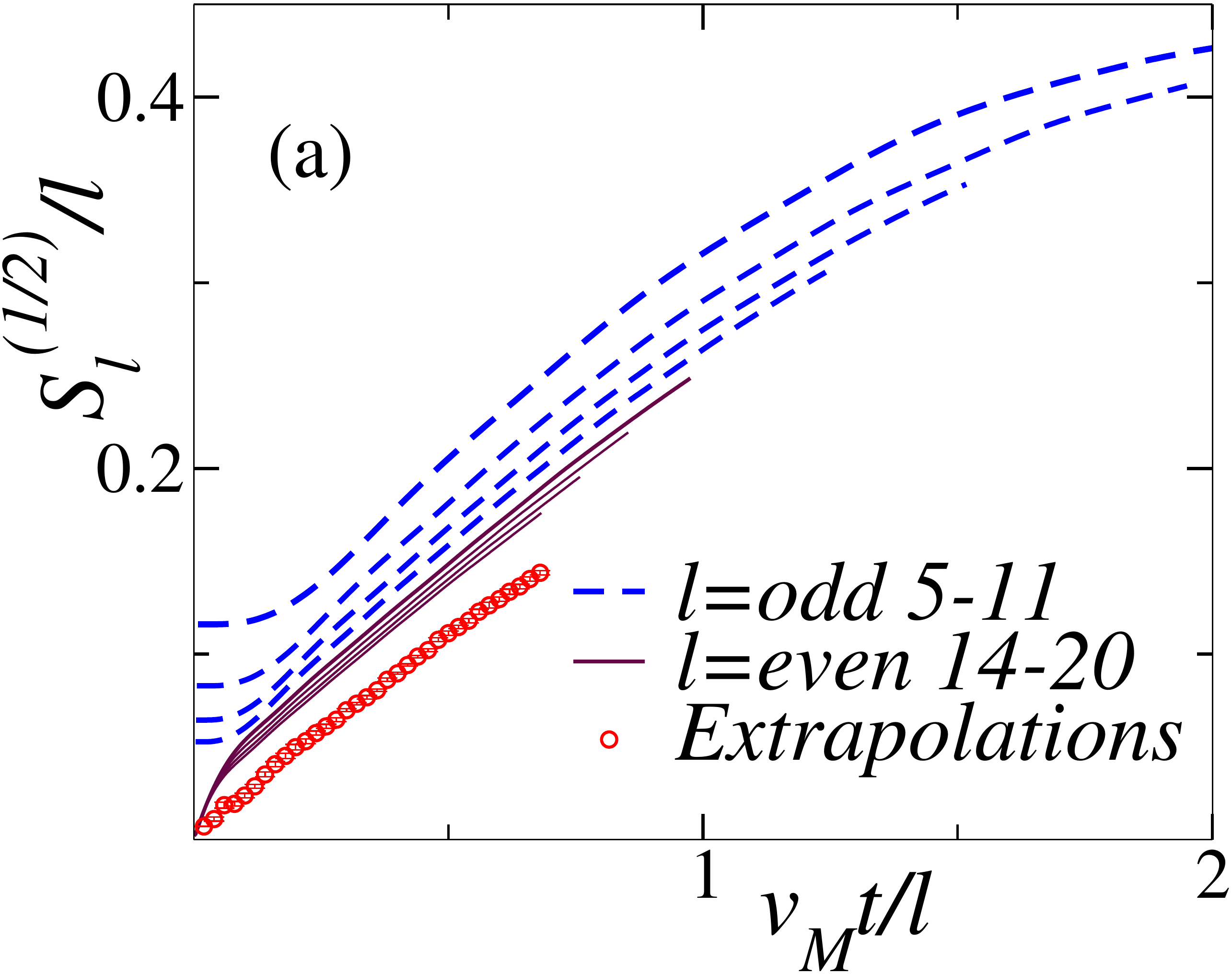} 
	\includegraphics[scale=0.22]{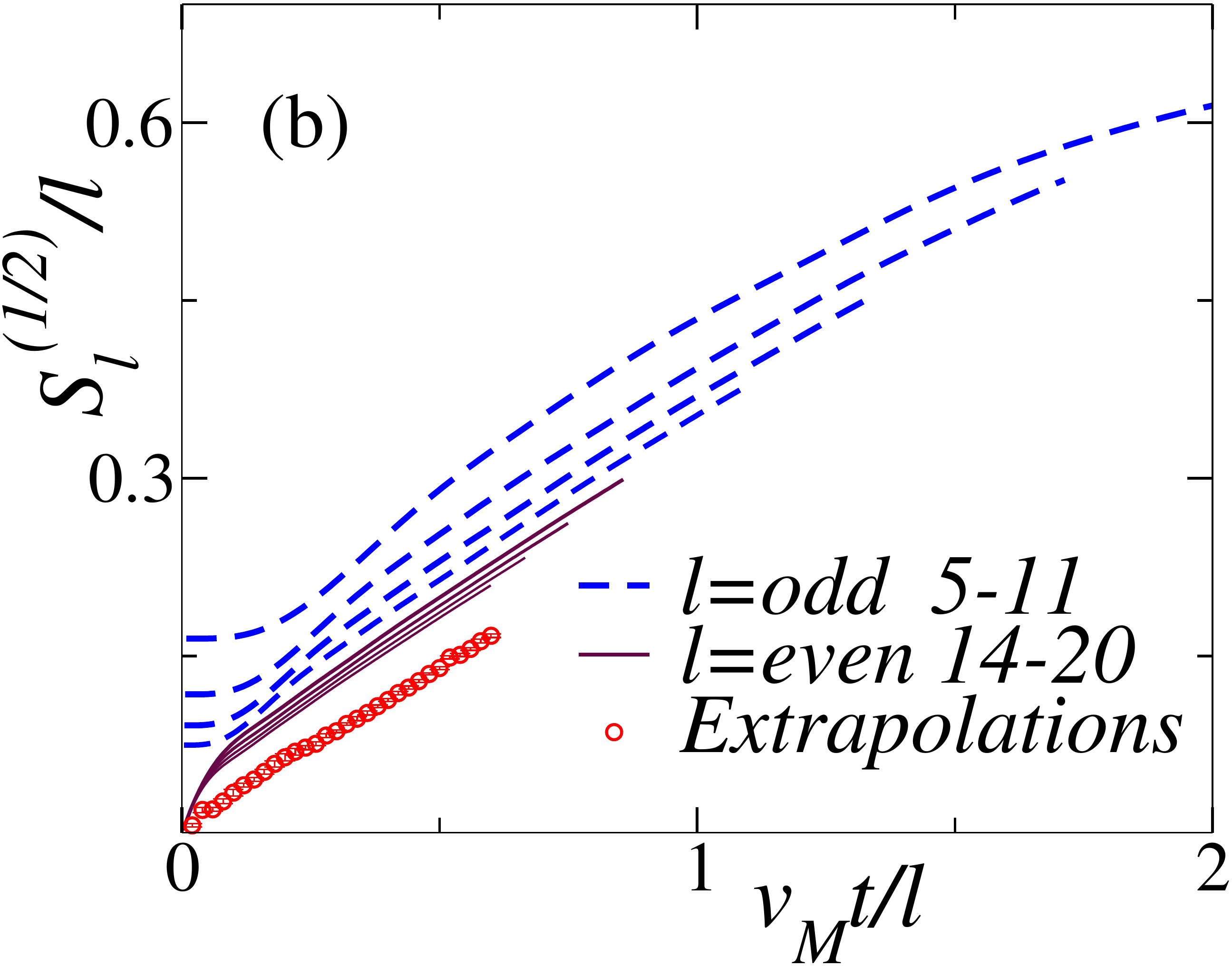} 
	\includegraphics[scale=0.22]{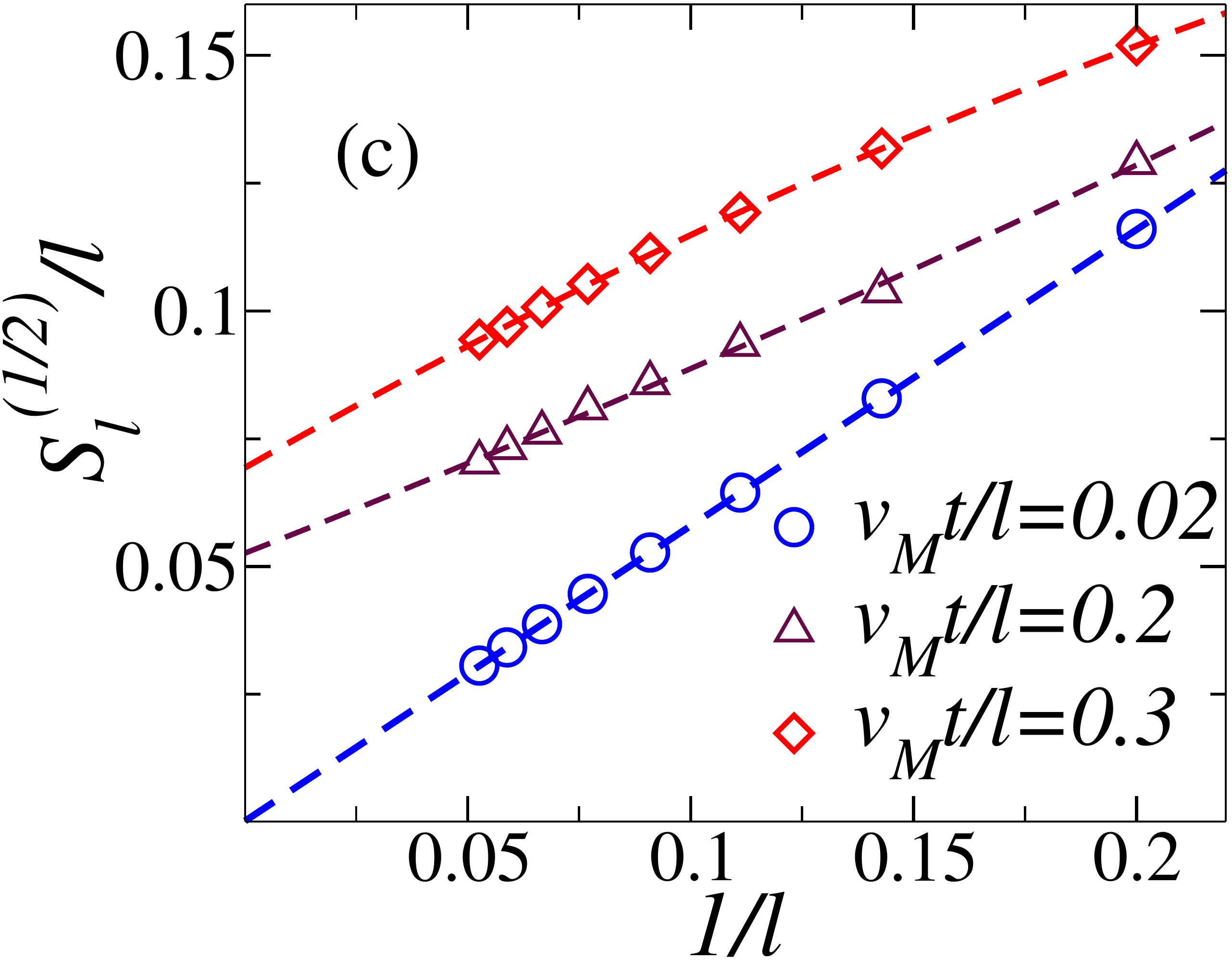} 
	\caption{R\`enyi entropy density  $S_l^{(1/2)}/l$ after the quench from integrable initial states, in a chain of $L=48$ sites. Subfigures $(a)$, $(b)$: same conventions as in Fig.~\ref{fig:vonNeumann}, for the initial states corresponding to the choices $c_{11}=0.95$, $c_{22}=0.3$, $c_{33}\simeq 0.087$ and $c_{11}=0.9$, $c_{22}=0.35$, $c_{33}\simeq 0.26$. Subfigure~$(c)$:  numerical extrapolations of the tDMRG results for finite odd values of $l$ (symbols). The dashed lines correspond to the quadratic curve $S^{(1/2)}_l/l=s_{\infty}+a/l+b/l^{2}$, where $s_{\infty}$, $a$, $b$ are fitting parameters.
	}
	\label{fig:Renyie}
\end{figure}

As we have mentioned in Sec.~\ref{sec:entanglement_dynamics}, we used the same extrapolation procedure to compute the dynamics of the R\'enyi entropy of order $\alpha$ in the scaling limit $t,l\to\infty$. An example of our result, corresponding to $\alpha=1/2$, is displayed in Fig.~\ref{fig:Renyie}. From the figure, we see that, as expected, the qualitative features of the R\'enyi and von Neumann entropies are similar, at least at the time scales available to our numerical procedure. 
We chose to report the R\'enyi entropy of order $1/2$ because it is related to the time evolution of the entanglement negativity \cite{ctc-14,AlCa19}, 
a relevant entanglement measure for tripartite systems.

\begin{thebibliography}{99}



\bibitem{CaEM16} 
P. Calabrese, F. H. L. Essler, and G. Mussardo, 
\href{http://dx.doi.org/10.1088/1742-5468/2016/06/064001}{J. Stat. Mech. (2016) 064001}.

\bibitem{KiWW06} 
T. Kinoshita, T. Wenger, and D. S. Weiss, 
\href{http://dx.doi.org/10.1038/nature04693}{Nature {\bf 440}, 900 (2006)}.

\bibitem{FSEH13} 
T. Fukuhara, P. Schauß, M. Endres, S. Hild, M. Cheneau, I. Bloch, and C. Gross, 
\href{http://dx.doi.org/10.1038/nature12541}{Nature {\bf 502}, 76 (2013)}.

\bibitem{LaGS16} 
T. Langen, T. Gasenzer, and J. Schmiedmayer, 
\href{http://dx.doi.org/10.1088/1742-5468/2016/06/064009}{J. Stat. Mech. (2016) 064009}.

\bibitem{LEGR15} 
T. Langen, S. Erne, R. Geiger, B. Rauer, T. Schweigler, M. Kuhnert, W. Rohringer, I. E. Mazets, T. Gasenzer, and J. Schmiedmayer, 
\href{http://dx.doi.org/10.1126/science.1257026}{Science {\bf 348}, 207 (2015)}.

\bibitem{SBDD19} 
M. Schemmer, I. Bouchoule, B. Doyon, and J. Dubail, 
\href{http://dx.doi.org/10.1103/PhysRevLett.122.090601}{Phys. Rev. Lett. {\bf 122}, 090601 (2019)}.


\bibitem{cc-06}
P. Calabrese and J. Cardy, 
\href{http://dx.doi.org/10.1103/PhysRevLett.96.136801}{Phys. Rev. Lett. {\bf 96}, 136801 (2006)};\\
P. Calabrese and J. Cardy, 
\href{http://dx.doi.org/10.1088/1742-5468/2007/06/P06008}{J. Stat. Mech. (2007) P06008}.



\bibitem{EsFa16} 
F. H. L. Essler and M. Fagotti, 
\href{http://dx.doi.org/10.1088/1742-5468/2016/06/064002}{J. Stat. Mech. (2016) 064002}.

\bibitem{Caux16} 
J.-S. Caux, 
\href{http://dx.doi.org/10.1088/1742-5468/2016/06/064006}{J. Stat. Mech. (2016) 064006}.

\bibitem{IMPZ16} 
E. Ilievski, M. Medenjak, T. Prosen, and L. Zadnik, 
\href{http://dx.doi.org/10.1088/1742-5468/2016/06/064008}{J. Stat. Mech. (2016) 064008}.


\bibitem{CaEs13} 
J.-S. Caux and F. H. L. Essler, 
\href{http://dx.doi.org/10.1103/PhysRevLett.110.257203}{Phys. Rev. Lett. {\bf 110}, 257203 (2013)}.


\bibitem{DWBC14} 
J. De Nardis, B. Wouters, M. Brockmann, and J.-S. Caux, 
\href{http://dx.doi.org/10.1103/PhysRevA.89.033601}{Phys. Rev. A {\bf 89}, 033601 (2014)}.


\bibitem{BWFD14} 
M. Brockmann, B. Wouters, D. Fioretto, J. De Nardis, R. Vlijm, and J.-S. Caux, 
\href{http://dx.doi.org/10.1088/1742-5468/2014/12/P12009}{J. Stat. Mech. (2014) P12009};\\
B. Wouters, J. De Nardis, M. Brockmann, D. Fioretto, M. Rigol, and J.-S. Caux, 
\href{http://dx.doi.org/10.1103/PhysRevLett.113.117202}{Phys. Rev. Lett. {\bf 113}, 117202 (2014)}.

\bibitem{PMWK14} 
B. Pozsgay, M. Mesty\'an, M. A. Werner, M. Kormos, G. Zar\'and, and G. Tak\'acs, 
\href{http://dx.doi.org/10.1103/PhysRevLett.113.117203}{Phys. Rev. Lett. {\bf 113}, 117203 (2014)};\\
M. Mesty\'an, B. Pozsgay, G. Tak\'acs, and M. A. Werner, 
\href{http://dx.doi.org/10.1088/1742-5468/2015/04/P04001}{J. Stat. Mech. (2015) P04001}.

\bibitem{IDWC15} 
E. Ilievski, J. De Nardis, B. Wouters, J.-S. Caux, F. H. L. Essler, and T. Prosen, 
\href{http://dx.doi.org/10.1103/PhysRevLett.115.157201}{Phys. Rev. Lett. {\bf 115}, 157201 (2015)}.

\bibitem{IQDB16} 
E. Ilievski, E. Quinn, J. De Nardis, and M. Brockmann, 
\href{http://dx.doi.org/10.1088/1742-5468/2016/06/063101}{J. Stat. Mech. (2016) 063101}.

\bibitem{PiVC16} 
L. Piroli, E. Vernier, and P. Calabrese, 
\href{http://dx.doi.org/10.1103/PhysRevB.94.054313}{Phys. Rev. B {\bf 94}, 054313 (2016)}.

\bibitem{PoVW17} 
B. Pozsgay, E. Vernier, and M. A. Werner, 
\href{http://dx.doi.org/10.1088/1742-5468/aa82c1}{J. Stat. Mech. (2017) 093103}.

\bibitem{IlQC17} 
E. Ilievski, E. Quinn, and J.-S. Caux, 
\href{http://dx.doi.org/10.1103/PhysRevB.95.115128}{Phys. Rev. B {\bf 95}, 115128 (2017)}.

\bibitem{PVCR17} 
L. Piroli, E. Vernier, P. Calabrese, and M. Rigol, 
\href{http://dx.doi.org/10.1103/PhysRevB.95.054308}{Phys. Rev. B {\bf 95}, 054308 (2017)}.



\bibitem{RDYO07} 
M. Rigol, V. Dunjko, V. Yurovsky, and M. Olshanii, 
\href{http://dx.doi.org/10.1103/PhysRevLett.98.050405}{Phys. Rev. Lett. {\bf 98}, 050405 (2007)}.

\bibitem{ViRi16} 
L. Vidmar and M. Rigol, 
\href{http://dx.doi.org/10.1088/1742-5468/2016/06/064007}{J. Stat. Mech. (2016) 064007}.

\bibitem{CaEF11} 
P. Calabrese, F. H. L. Essler, and M. Fagotti, 
\href{http://dx.doi.org/10.1103/PhysRevLett.106.227203}{Phys. Rev. Lett. {\bf 106}, 227203 (2011)};\\
P. Calabrese, F. H. L. Essler, and M. Fagotti, 
\href{http://dx.doi.org/10.1088/1742-5468/2012/07/P07016}{J. Stat. Mech. (2012) P07016};\\
P. Calabrese, F. H. L. Essler, and M. Fagotti, 
\href{http://dx.doi.org/10.1088/1742-5468/2012/07/P07022}{J. Stat. Mech. (2012) P07022}.

\bibitem{FaEs13} 
M. Fagotti and F. H. L. Essler, 
\href{http://dx.doi.org/10.1088/1742-5468/2013/07/P07012}{J. Stat. Mech. (2013) P07012}.

\bibitem{Pozs13} 
B. Pozsgay, 
\href{http://dx.doi.org/10.1088/1742-5468/2013/07/P07003}{J. Stat. Mech. (2013) P07003}.

\bibitem{Pros11} 
T. Prosen, 
\href{http://dx.doi.org/10.1103/PhysRevLett.106.217206}{Phys. Rev. Lett. {\bf 106}, 217206 (2011)}.

\bibitem{PrIl13} 
T. Prosen and E. Ilievski, 
\href{http://dx.doi.org/10.1103/PhysRevLett.111.057203}{Phys. Rev. Lett. {\bf 111}, 057203 (2013)}.

\bibitem{IlMP15} 
E. Ilievski, M. Medenjak, and T. Prosen, 
\href{http://dx.doi.org/10.1103/PhysRevLett.115.120601}{Phys. Rev. Lett. {\bf 115}, 120601 (2015)}.

\bibitem{PiVe16} 
L. Piroli and E. Vernier, 
\href{http://dx.doi.org/10.1088/1742-5468/2016/05/053106}{J. Stat. Mech. (2016) 053106}.

\bibitem{DeCD17} 
A. De Luca, M. Collura, and J. De Nardis, 
\href{http://dx.doi.org/10.1103/PhysRevB.96.020403}{Phys. Rev. B {\bf 96}, 020403 (2017)}.


\bibitem{BoEL14} 
L. Bonnes, F. H. L. Essler, and A. M. L\"auchli, 
\href{http://dx.doi.org/10.1103/PhysRevLett.113.187203}{Phys. Rev. Lett. {\bf 113}, 187203 (2014)}.

\bibitem{Pozs11} 
B. Pozsgay, 
\href{http://dx.doi.org/10.1088/1742-5468/2011/11/P11017}{J. Stat. Mech. (2011) P11017}.

\bibitem{MePo14} 
M. Mesty\'an and B. Pozsgay, 
\href{http://dx.doi.org/10.1088/1742-5468/2014/09/P09020}{J. Stat. Mech. (2014) P09020}.

\bibitem{NeSm13} 
S. Negro and F. Smirnov, 
\href{http://dx.doi.org/10.1016/j.nuclphysb.2013.06.023}{Nucl. Phys. B {\bf 875}, 166 (2013)},\\
S. Negro, 
\href{http://dx.doi.org/10.1142/S0217751X14501115}{Int. J. Mod. Phys. A {\bf 29}, 1450111 (2014)}.

\bibitem{BePC16} 
B. Bertini, L. Piroli, and P. Calabrese, 
\href{http://dx.doi.org/10.1088/1742-5468/2016/06/063102}{J. Stat. Mech. (2016) 063102}.

\bibitem{Pozs17} 
B. Pozsgay, 
\href{http://dx.doi.org/10.1088/1751-8121/aa5344}{J. Phys. A: Math. Theor. {\bf 50}, 074006 (2017)}.

\bibitem{BaPi18} 
A. Bastianello and L. Piroli, 
\href{http://dx.doi.org/10.1088/1742-5468/aaeb48}{J. Stat. Mech. (2018) 113104};\\
A. Bastianello, L. Piroli, and P. Calabrese, 
\href{http://dx.doi.org/10.1103/PhysRevLett.120.190601}{Phys. Rev. Lett. {\bf 120}, 190601 (2018)}.

\bibitem{AlCa17} 
V. Alba and P. Calabrese, 
\href{http://dx.doi.org/10.1073/pnas.1703516114}{PNAS {\bf 114}, 7947 (2017)}.

\bibitem{AlCa18} 
V. Alba and P. Calabrese, 
\href{http://dx.doi.org/10.21468/SciPostPhys.4.3.017}{SciPost Physics {\bf 4}, 017 (2018)}.

\bibitem{CaCa05} 
P. Calabrese and J. Cardy, 
\href{http://dx.doi.org/10.1088/1742-5468/2005/04/P04010}{J. Stat. Mech. (2005) P04010}.


\bibitem{KoPo12} 
K. K. Kozlowski and B. Pozsgay, 
\href{http://dx.doi.org/10.1088/1742-5468/2012/05/P05021}{J. Stat. Mech. (2012) P05021}.

\bibitem{Pozs14} 
B. Pozsgay, 
\href{http://dx.doi.org/10.1088/1742-5468/2014/06/P06011}{J. Stat. Mech. (2014) P06011}.

\bibitem{cd-14}
P. Calabrese and P. Le Doussal, 
\href{http://dx.doi.org/10.1088/1742-5468/2014/05/P05004}{J. Stat. Mech. (2014) P05004}.

\bibitem{PiCa14} 
L. Piroli and P. Calabrese, 
\href{http://dx.doi.org/10.1088/1751-8113/47/38/385003}{J. Phys. A: Math. Theor. {\bf 47}, 385003 (2014)}.

\bibitem{Broc14} 
M. Brockmann, 
\href{http://dx.doi.org/10.1088/1742-5468/2014/05/P05006}{J. Stat. Mech. (2014) P05006};\\
M. Brockmann, J. De Nardis, B. Wouters, and J.-S. Caux, 
\href{http://dx.doi.org/10.1088/1751-8113/47/34/345003}{J. Phys. A: Math. Theor. {\bf 47}, 345003 (2014)};\\
M. Brockmann, J. D. Nardis, B. Wouters, and J.-S. Caux, 
\href{http://dx.doi.org/10.1088/1751-8113/47/14/145003}{J. Phys. A: Math. Theor. {\bf 47}, 145003 (2014)}.

\bibitem{LeKZ15} 
M. de Leeuw, C. Kristjansen, and K. Zarembo, 
\href{http://dx.doi.org/10.1007/JHEP08(2015)098}{JHEP 98 (2015)};\\
I. Buhl-Mortensen, M. de Leeuw, C. Kristjansen, and K. Zarembo, 
\href{http://dx.doi.org/10.1007/JHEP02(2016)052}{JHEP 52 (2016)};\\
O. Foda and K. Zarembo, 
\href{http://dx.doi.org/10.1088/1742-5468/2016/02/023107}{J. Stat. Mech. (2016) 023107}.

\bibitem{LeKM16} 
M. de Leeuw, C. Kristjansen, and S. Mori, 
\href{http://dx.doi.org/10.1016/j.physletb.2016.10.044}{Phys. Lett. B {\bf 763}, 197 (2016)}.

\bibitem{HoST16} 
D. X. Horv\'ath, S. Sotiriadis, and G. Tak\'acs, 
\href{http://dx.doi.org/10.1016/j.nuclphysb.2015.11.025}{Nucl. Phys. B {\bf 902}, 508 (2016)};\\
D. X. Horv\'ath and G. Tak\'acs, 
\href{http://dx.doi.org/10.1016/j.physletb.2017.05.087}{Phys. Lett. B {\bf 771}, 539 (2017)};\\
D. X. Horv\'ath, M. Kormos, and G. Tak\'acs, 
\href{http://dx.doi.org/10.1007/JHEP08(2018)170}{JHEP 08 (2018) 170}.

\bibitem{BrSt17} 
M. Brockmann and J.-M. St\'ephan, 
\href{http://dx.doi.org/10.1088/1751-8121/aa809c}{J. Phys. A: Math. Theor. {\bf 50}, 354001 (2017)}.

\bibitem{Pozs18} 
B. Pozsgay, 
\href{http://dx.doi.org/10.1088/1742-5468/aabbe1}{J. Stat. Mech. (2018) 053103}.

\bibitem{LeKL18} 
M. de Leeuw, C. Kristjansen, and G. Linardopoulos, 
\href{http://dx.doi.org/10.1016/j.physletb.2018.03.083}{Phys. Lett. B {\bf 781}, 238 (2018)}.

\bibitem{HoKT19} 
K. H\'ods\'agi, M. Kormos, and G. Tak\'acs, 
\href{http://arxiv.org/abs/1905.05623}{arXiv:1905.05623 (2019)}.


\bibitem{PiPV17} 
L. Piroli, B. Pozsgay, and E. Vernier, 
\href{http://dx.doi.org/10.1088/1742-5468/aa5d1e}{J. Stat. Mech. (2017) 023106}.

\bibitem{PiPV17_II} 
L. Piroli, B. Pozsgay, and E. Vernier, 
\href{http://dx.doi.org/10.1016/j.nuclphysb.2017.10.012}{Nucl. Phys. B {\bf 925}, 362 (2017)}.

\bibitem{PiPV18_losch} 
L. Piroli, B. Pozsgay, and E. Vernier, 
\href{http://dx.doi.org/10.1016/j.nuclphysb.2018.06.015}{Nucl. Phys. B {\bf 933}, 454 (2018)}.

\bibitem{PoPV18} 
B. Pozsgay, L. Piroli, and E. Vernier, 
\href{http://dx.doi.org/10.21468/SciPostPhys.6.5.062}{SciPost Phys. {\bf 6}, 062 (2019)}.


\bibitem{Ghos94} 
S. Ghoshal, 
\href{http://dx.doi.org/10.1142/S0217751X94001941}{Int. J. Mod. Phys. A {\bf 09}, 4801 (1994)};\\
S. Ghoshal and A. Zamolodchikov, 
\href{http://dx.doi.org/10.1142/S0217751X94001552}{Int. J. Mod. Phys. A {\bf 09}, 3841 (1994)}.

\bibitem{Delf14} 
G. Delfino, 
\href{http://dx.doi.org/10.1088/1751-8113/47/40/402001}{J. Phys. A: Math. Theor. {\bf 47}, 402001 (2014)};\\
G. Delfino and J. Viti, 
\href{http://dx.doi.org/10.1088/1751-8121/aa5660}{J. Phys. A: Math. Theor. {\bf 50}, 084004 (2017)}.

\bibitem{Schu15} 
D. Schuricht, 
\href{http://dx.doi.org/10.1088/1742-5468/2015/11/P11004}{J. Stat. Mech. (2015) P11004}.

\bibitem{BeSE14} 
B. Bertini, D. Schuricht, and F. H. L. Essler, 
\href{http://dx.doi.org/10.1088/1742-5468/2014/10/P10035}{J. Stat. Mech. (2014) P10035}.

\bibitem{efgk-05}
F. H. L. Essler, H. Frahm, F. G\"ohmann, A. Kl\"umper, and V. E. Korepin,  
{\it The One-Dimensional Hubbard Model}, Cambridge University Press (2005).

\bibitem{BlDZ08} 
I. Bloch, J. Dalibard, and W. Zwerger, 
\href{http://dx.doi.org/10.1103/RevModPhys.80.885}{Rev. Mod. Phys. {\bf 80}, 885 (2008)}.

\bibitem{guan2013} 
X.-W. Guan, M. T. Batchelor, and C. Lee, 
\href{http://dx.doi.org/10.1103/RevModPhys.85.1633}{Rev. Mod. Phys. {\bf 85}, 1633 (2013)}.

\bibitem{PMCL14} 
G. Pagano, M. Mancini, G. Cappellini, P. Lombardi, F. Schafer, H. Hu, X.-J. Liu, J. Catani, C. Sias, M. Inguscio, and L. Fallani, 
\href{http://dx.doi.org/10.1038/nphys2878}{Nature Phys. {\bf 10}, 198 (2014)}.

\bibitem{exp2}
J. Vijayan, P. Sompet, G. Salomon, J. Koepsell, S. Hirthe, A. Bohrdt, F. Grusdt, I. Bloch, and C Gross,
\href{https://arxiv.org/pdf/1905.13638.pdf}{ArXiv:1905.13638}.


\bibitem{lai-74}
C. K. Lai, 
\href{http://dx.doi.org/10.1063/1.1666522}{J. Math. Phys. {\bf 15}, 1675 (1974)}.

\bibitem{sutherland-75}
B. Sutherland, 
\href{https://doi.org/10.1103/PhysRevB.12.3795}{Phys. Rev. B {\bf 12}, 3795 (1975)}.

\bibitem{PeSc81} 
J. H. H. Perk and C. L. Schultz, 
\href{http://dx.doi.org/10.1016/0375-9601(81)90994-4}{Phys. Lett. A {\bf 84}, 407 (1981)};\\
J.H.H. Perk and C.L. Schultz, 
in: Non-linear integrable systems -- classical theory and quantum theory,
Proceedings of RIMS Symposium organized by M. Sato, Kyoto, Japan,
13-16 May 1981, eds. M. Jimbo and T. Miwa,
(World Scientific, Singapore, 1983), pp. 135-152. 


\bibitem{Schu83} 
C. L. Schultz, 
\href{http://dx.doi.org/10.1016/0378-4371(83)90083-3}{Physica A: Stat. Mech. App. {\bf 122}, 71 (1983)}.


\bibitem{MBPC17} 
M. Mesty\'an, B. Bertini, L. Piroli, and P. Calabrese, 
\href{http://dx.doi.org/10.1088/1742-5468/aa7df0}{J. Stat. Mech. (2017) 083103}.

\bibitem{PVCP18_I} 
L. Piroli, E. Vernier, P. Calabrese, and B. Pozsgay, 
\href{http://dx.doi.org/10.1088/1742-5468/ab1c51}{J. Stat. Mech. (2019) 063103}.

\bibitem{PVCP18_II} 
L. Piroli, E. Vernier, P. Calabrese, and B. Pozsgay, 
\href{http://dx.doi.org/10.1088/1742-5468/ab1c52}{J. Stat. Mech. (2019) 063104}.



\bibitem{WhFe04} 
S. R. White and A. E. Feiguin, 
\href{http://dx.doi.org/10.1103/PhysRevLett.93.076401}{Phys. Rev. Lett. {\bf 93}, 076401 (2004)}.

\bibitem{DKSV04} 
A. J. Daley, C. Kollath, U. Schollw\"ock, and G. Vidal, 
\href{http://dx.doi.org/10.1088/1742-5468/2004/04/P04005}{J. Stat. Mech. (2004) P04005}.

\bibitem{Vida07} 
G. Vidal, 
\href{http://dx.doi.org/10.1103/PhysRevLett.98.070201}{Phys. Rev. Lett. {\bf 98}, 070201 (2007)}.

\bibitem{FCEC14} 
M. Fagotti, M. Collura, F. H. L. Essler, and P. Calabrese, 
\href{http://dx.doi.org/10.1103/PhysRevB.89.125101}{Phys. Rev. B {\bf 89}, 125101 (2014)}.

\bibitem{AlCal-b}
V.~Alba and P.~Calabrese, 
\href{https://doi.org/10.1103/PhysRevB.96.115421}{Phys.\ Rev.\ B {\bf 96}, 115421 (2017)};\\
V. Alba and P. Calabrese, 
\href{http://dx.doi.org/10.1088/1742-5468/aa934c}{J. Stat. Mech. (2017) 113105};\\
M.~Mesty\'an, V.~Alba, and P.~Calabrese, 
\href{https://doi.org/10.1088/1742-5468/aad6b9}{J. Stat. Mech. (2018) 083104}. 

\bibitem{AlCal-c}
V.~Alba and P.~Calabrese, 
\href{https://arxiv.org/abs/1809.09119}{arXiv:1809.09119}. 

\bibitem{BeTC17} 
B. Bertini, E. Tartaglia, and P. Calabrese, 
\href{http://dx.doi.org/10.1088/1742-5468/aa8c2c}{J. Stat. Mech. (2017) 103107}.

\bibitem{BeTC18} 
B. Bertini, E. Tartaglia, and P. Calabrese, 
\href{http://dx.doi.org/10.1088/1742-5468/aac73f}{J. Stat. Mech. (2018) 063104}.

\bibitem{ZhVR18} 
Y. Zhang, L. Vidmar, and M. Rigol, 
\href{http://dx.doi.org/10.1103/PhysRevA.98.042129}{Phys. Rev. A {\bf 98}, 042129 (2018)}.

\bibitem{ZhVR19} 
Y. Zhang, L. Vidmar, and M. Rigol, 
\href{http://arxiv.org/abs/1903.10521}{arXiv:1903.10521 (2019)}.




\bibitem{kr-81}
P. P. Kulish and N. Y. Reshetikhin, 
Sov. Phys. JETP {\bf 53} (1981)

\bibitem{johannesson-86}
H. Johannesson, 
\href{https://doi.org/10.1016/0375-9601(86)90300-2}{Phys. Lett. A {\bf 116}, 133 (1986)}.

\bibitem{johannesson2-86}
H. Johannesson, 
\href{https://doi.org/10.1016/0550-3213(86)90554-7}{Nucl. Phys. B {\bf 270}, 235 (1986)}.

\bibitem{kbi-93} V.E. Korepin, N.M. Bogoliubov and A.G. Izergin, 
{\it Quantum inverse scattering method and correlation functions}, Cambridge University Press (1993).

\bibitem{MBPC19} 
M. Mesty\'an, B. Bertini, L. Piroli, and P. Calabrese, 
\href{http://dx.doi.org/10.1103/PhysRevB.99.014305}{Phys. Rev. B {\bf 99}, 014305 (2019)}.

\bibitem{takahashi_book}
M. Takahashi, Thermodynamics of One-Dimensional Solvable Models (Cambridge University Press, Cambridge, UK, 1999).

\bibitem{PVWC06} 
D. Perez-Garcia, F. Verstraete, M. M. Wolf, and J. I. Cirac, 
Quantum Inf. Comput. {\bf 7}, 401 (2007).

\bibitem{YaYa69} 
C. N. Yang and C. P. Yang, 
\href{http://dx.doi.org/10.1063/1.1664947}{J. Math. Phys. {\bf 10}, 1115 (1969)}.

\bibitem{Al-18}
V.~Alba, 
\href{https://doi.org/10.1103/PhysRevB.97.245135}{Phys. Rev. B {\bf 97}, 245135 (2018)}.

\bibitem{bertini-2018a}
B.~Bertini, M.~Fagotti, L.~Piroli, and P.~Calabrese, 
\href{https://doi.org/10.1088/1751-8121/aad82e}{J.\ Phys.\ A  {\bf 51}, 39LT01 (2018)}. 

\bibitem{abf-19}
V.~Alba, B.~Bertini, and M.~Fagotti, 
\href{https://arxiv.org/pdf/1903.00467.pdf}{arXiv:1903.00467}. 

\bibitem{CaDY16} 
O. A. Castro-Alvaredo, B. Doyon, and T. Yoshimura, 
\href{http://dx.doi.org/10.1103/PhysRevX.6.041065}{Phys. Rev. X {\bf 6}, 41065 (2016)}.


\bibitem{BCDF16} 
B.~Bertini, M.~Collura, J.~De~Nardis, and M.~Fagotti, 
\href{http://dx.doi.org/10.1103/PhysRevLett.117.207201}{Phys. Rev. Lett. {\bf 117}, 207201 (2016)}.

\bibitem{bc-18}
A. Bastianello and P. Calabrese, 
\href{https://doi.org/10.21468/SciPostPhys.5.4.033}{SciPost Phys. {\bf 5}, 033 (2018)}.


\bibitem{asplund}
C.~T.~Asplund and A.~Bernamonti, 
\href{https://doi.org/10.1103/PhysRevD.89.066015}{Phys.\ Rev.\ D {\bf 89}, 066015 (2014)};\\
V.~Balasubramanian, A.~Bernamonti, N.~Copland, B.~Craps, and F.~Galli, 
\href{https://doi.org/10.1103/PhysRevD.84.105017}{Phys.\ Rev.\ D {\bf 84}, 105017 (2011)};\\
C.~T.~Asplund, A.~Bernamonti, F.~Galli, and T.~Hartmann, 
\href{http://dx.doi.org/10.1007/JHEP09(2015)110}{JHEP 09 (2015) 110}.

\bibitem{AlCa19} 
V. Alba and P. Calabrese, 
\href{http://arxiv.org/abs/1903.09176}{arXiv:1903.09176 (2019)}.

\bibitem{RiKl19} 
G. A. P. Ribeiro and A. Klümper, 
\href{http://dx.doi.org/10.1088/1742-5468/aaf31e}{J. Stat. Mech. (2019) 013103}.



\bibitem{MWNM09} 
S. R. Manmana, S. Wessel, R. M. Noack, and A. Muramatsu, 
\href{http://dx.doi.org/10.1103/PhysRevB.79.155104}{Phys. Rev. B {\bf 79}, 155104 (2009)}.


\bibitem{itensor}
ITensor, Intelligent Tensor C ++ library 
\href{http://itensor.org/}{http://itensor.org}

\bibitem{ctc-14} 
A. Coser, E. Tonni, and P. Calabrese, 
\href{https://doi.org/10.1088/1742-5468/2014/12/P12017}{J. Stat. Mech.  (2014) P12017}.

\end{thebibliography}
\end{document}